\documentclass[a4paper,11pt]{article}
\usepackage{jheppub}
\usepackage{dsfont}
\usepackage{amsmath,amssymb,amscd,amsfonts,mathtools}

\usepackage[toc,page]{appendix}
\usepackage{epsfig}
\usepackage{epstopdf}
\usepackage{latexsym}
\usepackage{graphicx}
\usepackage{placeins}
\usepackage{floatrow}
\usepackage{subfig}
\usepackage{caption}
\usepackage{booktabs}
\usepackage{bbm}
\usepackage{color,soul}
\usepackage{physics}
\usepackage{stackrel}
\usepackage{tensor}
\usepackage{tikz}
\usepackage{comment}
\usetikzlibrary{matrix}
\usetikzlibrary{decorations.markings,calc,shapes,decorations.pathmorphing,patterns,decorations.pathreplacing,arrows.meta}
\usetikzlibrary{positioning}
\usepackage{hyperref}

\pdfoutput=1
\makeatletter
\def\@fpheader{\relax}
\makeatother

\usepackage{wasysym}

\catcode`,\active

\catcode`\,12

\preprint{UT-WI-30-2025}
\title{Dissipation in Open Holography}

\author{Andreas Karch and Merna Youssef}

\affiliation{Theory Group, Weinberg Institute, Department of Physics, University of Texas,
2515 Speedway, Austin, TX 78712, USA.}

\emailAdd{karcha@utexas.edu}
\emailAdd{myoussef@utexas.edu}
\abstract{We exploit the holographic realization of a conformal theory coupled to an external bath realized via a double trace deformation and its gravity dual in terms of transparent boundary conditions in order to map out some basic dissipative properties of this simple open holographic system. In particular, we determine the energy transmission coefficient across the boundary, discover a novel duality relating weak and strong coupling to the external bath, and quantify the dissipation in the system by working out the quasi normal modes.} 

\begin{document}
\maketitle
\section{Introduction}

Understanding mixed phases and open systems is one of the cutting edge topics in contemporary quantum physics. While by now a classification of phases of matter based on groundstate properties is fairly well understood, classifying mixed states as they would result in any open system is far from complete. Definitions of mixed state phases have for example been put forward in \cite{Hastings:2011iig,Coser:2019ecy,Sang:2025ssd} and many interesting mixed states have been found partially based on these works. Coupling a system to an environment does not just wash out some of the features of the groundstate, it can realize completely new phases whose existence relies on the presence of dissipation \cite{Verstraete:2009zet,Diehl_2008,Fan_2024,Lee_2023,Zou_2023,Lu_2023,leefisher_2022,Zhu_2023,balasubramanian_2025}.

Open systems are also of immense practical importance. No quantum system exists in isolation. Usually dissipation is seen as the enemy, a problem to overcome in order to harness the power of quantum mechanics. What the recent advances make clear is that sometimes dissipation can also be used as a resource to realize physics that would be impossible without it.

Seemingly unrelated, open systems also played an important role in recent studies of quantum properties of black holes in AdS \cite{penington2020entanglementwedgereconstructioninformation,Almheiri_2019}. Holography in the form of the AdS/CFT correspondence provides a full UV complete definition of quantum gravity in asymptotically AdS spacetimes, making it the ideal laboratory to study questions such as black hole evaporation in a well controlled setting. Alas, at least for large black holes in AdS the reflecting boundary conditions of AdS prevent evaporation, so to allow it to happen one must couple the system to a thermal bath via modified boundary conditions \cite{Porrati_2002,Porrati_2003,Almheiri_20202,Geng:2020qvw,chen2020quantumextremalislandseasy,Geng:2020fxl}. Beyond being useful for studying black hole evaporation, this very same setting gives a simple example of a holographic open system. In this work, we will exploit this construction to study open holographic theories. Given the important insights holography has provided on the physics of generic pure strongly correlated quantum systems, we fully expect that it will also be able to do the same for open strongly correlated systems.

The simplest scenario to realize open holography along these lines is to couple the $d$ dimensional CFT$_d$ with it's AdS$_{d+1}$ dimensional gravitational dual to a second "bath" CFT$_d$, which itself may also have a dual description in terms of a different AdS$_{d+1}$, through a double-trace deformation.
From the bulk point of view this is implemented via transparent boundary conditions that allow the two AdS spaces to communicate with each other. 
This system, introduced in \cite{Aharony:2005sh,Aharony:2006hz,Kiritsis:2006hy,Karch:2023wui} based on the general treatment of multi-trace operators in holography worked out in \cite{Aharony:2001pa,Witten:2001ua},
will be our main tool for studying the dynamics of coupling a CFT with gravity dual to a bath. 

The idea to use holography to learn about open systems isn't entirely new. A recent example of exploiting exactly the same double trace construction to analyze open holographic system was given in \cite{Geng:2023ynk}, focusing mostly on a careful study of quantum aspects of the bulk physics. A similar system with a much simpler bath was studied in \cite{Rocha_2008}. In \cite{Jana:2020vyx,Loganayagam_2023} the authors also studied holographic systems coupled to other degrees of freedom in the context of the Schwinger Keldysh formalism, but mostly with reversed roles: the holographic system served as the bath. In \cite{ishii2025lindbladdynamicsholography} the authors gave a holographic implementation of the Lindblad formalism for open quantum systems where the bath degrees of freedom are effectively integrated out. In contrast, we will explicitly keep the full dynamics of the bath degrees of freedom. A somewhat orthogonal approach to open holography was also pursued in \cite{Pelliconi:2023ojb} and \cite{Arenas-Henriquez:2025rpt}.

To gain a first look at the dynamics of our doulbe trace coupled open system, we work out a few physically important properties: we quantify the transition coefficient of energy across the boundary, exhibit an interesting strong-weak coupling duality in the system, and then focus on the quasi-normal modes in this particular realization of a bath. Most notably we will verify that the coupling to the bath gives rise to imaginary frequencies even if the physical system is kept at zero temperature, allowing us to quantify the amount of dissipation induced by the bath. These studies should lay the foundation for any future investigations using this particular system as a workhorse to study open holography.

This paper is organized as follows. In section 2 we review the basic construction of two CFTs coupled via double traces and the corresponding bulk dual in terms of transparent boundary conditions. In section 3, we study energy flux and transmission coefficients across the boundary between the two AdS space. We find that the transmission coefficients are characteristics of the boundary conditions only, independent of the details of the incoming excitations, similar to what happens in 2d conformal interfaces \cite{Quella:2006de,Meineri:2019ycm}. In section 4, we analyze a new strong/weak duality due to the coupling: the full phase space of the coupling $g=h(2\Delta-d)$ is covered up to unity if we were to cover the whole regime of the conformal dimension. In other words, if we start with a black hole in the standard quantization coupled to an empty AdS in alternate quantization with coupling constant $g>1$ we get the modes of a black hole in the alternate quantization coupled to an empty AdS in the standard quantization. In section 5, we review quasi normal modes for empty AdS and BTZ black hole. Finally in section 6 and 7, we provide the numerical results for quasinormal modes and show the duality and resulting dissipation due to the coupling. We conclude with some exciting directions to investigate quantum open systems in holography. 

\section{Review of Double Trace Coupled Bath}
We start by recalling some standard knowledge about the asymptotics of the scalar field profile. The near boundary behavior of a scalar with mass $m$ is given by:
\begin{equation}
    \phi = \alpha \, z^{d-\Delta}+ \beta\, z^\Delta  + \ldots \qquad \text{as} \quad z\to 0
    \label{asymscalar}
\end{equation} 
where $\Delta_{\pm} = d/2 \pm  \sqrt{d^2+ 4 m^2}/2$.
The action for this massive scalar in $AdS_{d+1}$ is
\begin{eqnarray}
    S &=& -\frac{1}{2}\int d^{d+1}x\sqrt{g}\,(g^{\mu \nu}\partial_\mu \phi \, \partial_{\nu}\phi+m^2\phi^2) \nonumber  \\ &=& - \frac{1}{2} \int d^dx\,dz\, z^{-d+1}((\partial_z\phi)^2+(\partial_{\mu}\phi)^2+\frac{m^2}{z^2}\phi^2).
\end{eqnarray}

For $\tfrac{d}{2}<\Delta < d$, both terms vanish near the boundary, but the $\alpha$ term goes to zero more slowly and leads to a non-normalizable solution. We therefore fix $\alpha$ as the source and interpret $\beta$ as the VEV. This is called the standard quantization. In the regime $\tfrac{d}{2}<\Delta<\tfrac{d+2}{2}$ there is a second quantization scheme possible \cite{Breitenlohner:1982bm} which instead assigns the operator a dimension $\Delta$ with $\tfrac{d-2}{2}<\Delta<\tfrac{d}{2}$. For a scalar with mass in this window allowing both quantizations the dimension of alternate and standard scheme add up to $d$. Continuing to write the expansion of the scalar field as \eqref{asymscalar} even when $\Delta<d/2$, the $\alpha$ term now seems to go to zero faster, but in this alternate scheme it nevertheless continues to be the non-normalizable and hence the source term.

From the field-theory perspective, the bulk scalar field $\phi(z,t,\vec{x})$ in $AdS_{d+1}$ is dual to a single-trace operator $\mathcal{O}(t,\vec{x})$ in the boundary $CFT_d$. Fixing a finite value of $\alpha$ corresponds to setting boundary conditions for the source of $\mathcal{O}(t,\vec{x})$. The AdS/CFT dictionary then reads
\begin{equation}
    Z[J(t,\vec{x})]_{CFT} = \langle e^{i\int dt d^{d-1}\vec{x}\, J \, \mathcal{O}}\rangle_{CFT} = Z[J]_{AdS_{d+1}}
\end{equation}
where $J(t,\vec{x})$ is the source (i.e. the boundary condition) in the boundary CFT for the primary scalar operator—identified with $\alpha$ in the asymptotic expansion in the standard quantization. If we set $\alpha=0$, we are studying the undeformed CFT. If we instead set $\alpha \neq 0$, we obtain a deformed CFT with boundary  conditions $\alpha=J$:
\begin{equation}
    S_{CFT}\to S_{CFT}+\int d^dx\, \alpha(x)\mathcal{O}(x).
\end{equation}
In this work, we focus on a system of two originally decoupled CFTs, where one we think of as the physical system and the other as the bath, with a marginal deformation 
\begin{equation}
    S_{CFT}\to S_{CFT}+ h\,\int d^dx \, \mathcal{O}_L(x)\mathcal{O}_R(x)
\end{equation}
The operators $O_L$ and $O_R$ are scalar primaries from $CFT_L$ and $CFT_R$ respectively and so the double trace deformations couples the two systems to each other. For the interaction to be marginal we have to require
$\Delta_L+\Delta_R=d$. Note that this in particular implies that that the corresponding bulk scalar fields $\phi_L$ and $\phi_R$ have identical mass. Furthermore, if -say- the right CFT employs a scalar field in standard quantization (which is the choice we will make throughout this paper) the left will have to be in alternate (and vice versa).\footnote{The case $\Delta_L = \Delta_R = d/2$ is special in that it triggers a logarithmic running of the coupling and doesn't really lead to a marginal interaction \cite{Witten:2001ua} and we will not consider it further.} 

 On the bulk side, both $AdS_{d+1}$ spacetimes are glued together along the boundary, so that degrees of freedom can transfer from one to the other. In this case, the boundary conditions of AdS are no longer reflecting but instead become transparent boundary conditions:
\begin{equation}
\alpha_L(\vec x)= h\,(2 \Delta -d) \beta_R(\vec x), \quad \alpha_R(\vec x)= h\,(2 \Delta -d) \beta_L(\vec x)
\label{eq:mixingrelation}
\end{equation}
where subscripts denote fields in the first or second AdS space.

To understand the $(2 \Delta -d)$ pre-factor in the equation above, we take a step back to review the relation between the coefficient $\beta$ and vacuum expectation value in the case of a single CFT. For simplicity let us go to Euclidean signature for this was also done in the original \cite{Klebanov_1999}. A Lorentzian discussion can be found in \cite{Son:2002sd}. Also, let us focus on the standard quantized case, $\Delta > d/2$. The discussion of the alternate case requires some extra care but gives the same answer \cite{Klebanov_1999}. 
Solving the Euclidean Klein-Gordon equation for the scalar field with a given $\alpha(\vec{x})$ yields for the coefficient $\beta(\vec{x})$
\begin{equation}
\label{betais}
\beta(\vec{x})
= \pi^{-d/2} \frac{\Gamma(\Delta)}{\Gamma(\Delta-d/2)}\, \int d^d x' \frac{\alpha(\vec{x}')}{|\vec{x} - \vec{x}'|^{2 \Delta}}
\end{equation}
so that for the special case of a delta function source, $\alpha(\vec{x}) = \delta(\vec{x})$ one simply finds
\begin{equation}
\beta(\vec{x})
= \pi^{-d/2} \frac{\Gamma(\Delta)}{\Gamma(\Delta-d/2)}\, \frac{1}{|\vec{x} |^{2 \Delta}} 
\end{equation}
which indeed has the right functional form for a 1-pt function. To extract the normalization we need to compare to the  on-shell action which for the above solution evaluates to
\begin{equation}
I
= (\frac{d}{2} - \Delta) \pi^{-d/2} \frac{\Gamma(\Delta)}{\Gamma(\Delta-d/2)}\,\int d^d x \int d^d x' \frac{\alpha(\vec{x})\alpha(\vec{x}')}{|\vec{x} - \vec{x}'|^{2 \Delta}}.
\end{equation}
From this we calculate the two-point function
\begin{equation}
\langle O(\vec{x}) O(0) \rangle = (d - 2 \Delta) 
\pi^{-d/2} \frac{\Gamma(\Delta)}{\Gamma(\Delta-d/2)}\frac{1}{|\vec{x} |^{2 \Delta}}
\end{equation}
by varying with respect to the source $\alpha$ twice which indeed has the expected form. Varying with respect to $(-\alpha)$ only once gives the 1-pt function
\begin{equation}
\label{onepoint}
\langle O(\vec{x}) \rangle
= (2 \Delta - d) \pi^{-d/2} \frac{\Gamma(\Delta)}{\Gamma(\Delta-d/2)}\, \int d^d x' \frac{\alpha(\vec{x}')}{|\vec{x} - \vec{x}'|^{2 \Delta}}
\end{equation}
Comparing the exact expression \eqref{onepoint} with the formulate \eqref{betais} for $\beta$ we see that $\beta$ in fact is proportional to the one point function with the prefactor given by
\begin{equation}
\label{onepoint}
\langle O(\vec{x}) \rangle
= (2 \Delta - d) \beta(\vec{x}).
\end{equation}
How is this related to the boundary conditions
\eqref{eq:mixingrelation}? As far as the left CFT is concerned, the coupling constant, that is the coefficient of $O_L$ in the action, is $h \langle O_R \rangle$, whereas as far as the right CFT is concerned its coupling constant is $h \langle O_R \rangle$. Eq. \eqref{eq:mixingrelation} simply imposes this fact by demanding that the corresponding $\alpha$ coefficients are equal to these dynamically determined couplings, with the prefactor being inherited from the relation between 1-pt function and $\beta$ we just reviewed.

Let us end this review section with comments on potential top down realizations of our scenario, that is ways to embed them in a UV complete string theory with known dual CFTs. All that is required is to find two known top down examples of CFTs of the same $d$ whose bulk duals involve scalar fields of the same mass $m$, but one with alternate quantization. This is not entirely trivial, but examples can be found. Most famous AdS/CFT pairs, like the duality between ${\cal N}=4$ SYM and its AdS$_5$ $\times$ $S^5$ dual have no scalars with alternate quantizations. In $d=4$ these would be operators with dimension between 1 and 2, the smallest dimension operator in ${\cal N} =4$ is the dimension 2 scalar bi-linear, which correponds to the borderline $\Delta=d/2$ case that is not truly marginal \cite{Witten:2001ua}. One well-studied example of a CFT with known gravity dual that does, in fact, employs an alternate quantized scalar is the Klebanov-Witten CFT \cite{Klebanov:1998hh} which relies on a product gauge group with bi-fundamental matter and a quartic superpotential, also in $d=4$. Here, for the coupling to be marginal, the fundamental scalar fields have dimensions 3/4, so that the lowest dimension gauge invariant bi-linear scalar operator has dimension $3/2 < d/2$. We can choose this as our left CFT. One interesting aspect of the alternate quantization is that a double trace deformation of the form $O_R^2$ in this case is relevant and drives the theory to a new fixed point. This new fixed points is believed to be simply described by the theory with standard quantization. In the case of the Klebanov-Witten theory this deformed theory yields a second possible AdS/CFT pair based on the same supergravity solution, albeit with different boundary conditions, breaking all supersymmetry. With this we have all ingredients in place to realize our setting from the top down, where both left and right CFT can be chosen to be the Klebanov-Witten CFT, with the right side driven to a new fixed point by $O_R^2$ before coupling the two.

\section{Energy Flux and Transmission Coefficient}
\subsection{Setup}

As just reviewed, we consider two copies of AdS$_{d+1}$ with scalar fields $\phi_L$, $\phi_R$. For each one we have an expansion \eqref{asymscalar}
\begin{equation}
\phi(z,t) = \alpha(t) z^{d - \Delta} + \beta(t) z^{\Delta} + \dots
\label{asymscalartwo}
\end{equation}
$\alpha$ is the source and $\beta$ the vev in both standard  ($\Delta > d/2$) and alternate ($\Delta  < d/2)$ quantizations. We would like to calculate the transmission coefficient across the AdS boundary between the two regions implied by the transparent boundary contidions \eqref{eq:mixingrelation}. Unfortunately the asymptotic form \eqref{asymscalartwo} doesn't naturally lend itself to an interpretation in terms of incoming and outgoing waves. In order to extract transmission coefficients, one way to proceed is to tag solutions as ingoing and outgoing by considering the associated energy fluxes $T_t^z$.

In terms of the scalar the $T_t^z$ component of the stress tensor reads
\begin{equation}
T_{zt} = T_{zt}^s + k T_{zt}^i =  \nabla_z \phi \nabla_t \phi + \frac{k}{2} \nabla_z \nabla_t (\phi^2)
\end{equation}
where the first term is the standard stress tensor
\begin{equation}
T_{\mu \nu}^s = \partial_\mu \phi \partial_\nu \phi - \frac{1}{2} g_{\mu\nu} (\partial \phi)^2
- \frac{1}{2} g_{\mu\nu} m^2 \phi^2
\end{equation}
and the second with coefficient $k$ the allowed improvement term used by Breitenlohner and Freedman \cite{Breitenlohner:1982bm}
\begin{equation}
T_{\mu \nu}^i =  - \frac{k}{2} \left ( g_{\mu\nu} \Box - \nabla_\mu \nabla_\nu + R_{\mu\nu} \right) \phi^2
\end{equation}

Using
\begin{eqnarray}
\partial_z \phi &=& \Delta \beta z^{\Delta - 1} + (d - \Delta) \alpha z^{d - \Delta - 1} \\
\partial_t \phi &=& \dot{\beta} z^{\Delta} + \dot{\alpha} z^{d - \Delta} \\
\frac{1}{2} \nabla_z \nabla_t (\phi^2) &=&  \left[ \partial_z \partial_t (\phi^2) + \frac{1}{z} \partial_t (\phi^2) \right]
\end{eqnarray}
we get for the standard term
\begin{equation}
T_{zt}^s =
\Delta \beta \dot{\beta} z^{2\Delta - 1} +
\left( \Delta \beta \dot{\alpha} + (d - \Delta) \alpha \dot{\beta} \right) z^{d-1} +
(d - \Delta) \alpha \dot{\alpha} z^{2d - 2\Delta-1} .
\end{equation}
The improvement term evaluates to
\begin{equation}
T_{zt}^i =
(2 \Delta+1) \beta \dot{\beta} z^{2\Delta - 1} +
(d+1) \, \left(  \beta \dot{\alpha} +  \alpha \dot{\beta} \right) z^{d-1} +
(2 d - 2 \Delta +1) \alpha \dot{\alpha} z^{2 d - 2\Delta -1} .
\end{equation}

\noindent Last but not least let us define the energy flux density through a surface at $z=\epsilon$ with induced volume element $\sqrt{-g_I} = z^{-d}$ and unit normal $n_z =z^{-1}$:
\begin{equation}
\mathcal{F} = \sqrt{-g_I} \, T^z_t n_z = z^{-d+1} T_{zt}
\end{equation}
to find the corresponding contributions to the energy flux:
\begin{equation}
\label{improv}
\mathcal{F}^s =
\Delta \beta \dot{\beta} z^{2\Delta - d} +
\left( \Delta \beta \dot{\alpha} + (d - \Delta) \alpha \dot{\beta} \right)  +
(d - \Delta) \alpha \dot{\alpha} z^{d - 2\Delta} .
\end{equation}
In this form we can explicitly see that the last term diverges for standard quantization (which hence usually requires $\alpha=0$), while the first term diverges for alternate quantization. In the alternate case we are therefore forced to add the improvement term to ensure that also in this case setting the source to zero is the correct boundary condition. The middle term is finite as it is but vanishes with reflecting boundary conditions, both standard and alternate, and so there will be no flux across the boundary, as expected.
The improvement term evaluates to
\begin{equation}
\mathcal{F}^i =
(2 \Delta+1) \beta \dot{\beta} z^{2\Delta - d} + (d+1) \,
\left(  \beta \dot{\alpha} +  \alpha \dot{\beta} \right)  +
(2 d - 2 \Delta +1) \alpha \dot{\alpha} z^{d - 2\Delta } .
\end{equation}
For the special choice used in the BF paper
\begin{equation}
k = - \frac{\Delta}{2 \Delta +1}
\end{equation}
the $z^{2 \Delta -d}$ term cancels and so there is no divergence in the alternate quantization either.

\subsection{Transparent Boundary Condition}
\label{open}
We now want to apply these general insights to the transparent boundary conditions describing coupling the two CFTs dual to the two AdS space via a double trace deformation $\sim h O_L O_R$. The corresponding transparent boundary conditions are \eqref{eq:mixingrelation}:
\begin{equation}
\alpha_L = h (2 \Delta_R-d) \beta_R, \quad \alpha_R = h (2 \Delta_L -d) \beta_L \label{bcs}.
\end{equation}
For the double trace deformation to be marginal we want the two dimensions to obey
\begin{equation}
\Delta_R = d - \Delta_L
\end{equation}
which, as we stated before, also implies that if one of the two is in standard quantization, the other is in alternate. Without loss of generalization we take the left to be alternate ($\Delta_L < d/2$), the right to be standard ($\Delta_R > d/2$).
With this we get the following fluxes:

\vskip10pt
\noindent{\bf Alternate Quantization (Left):}
In order to ensure a finite flux on the left we can simply use the improvement term with $k_L=- \frac{\Delta_L}{2 \Delta_L +1}$. This cancels the divergent term and, dropping the term with positive powers of $z$ which vanishes near the boundary, the finite flux on the left reads
\begin{equation}
\mathcal{F}_{L} = \frac{d- 2 \Delta_L}{2 \Delta_L+1} \left ( \alpha_L \dot{\beta}_L (\Delta_L+1) - \Delta_L \dot{\alpha}_L \beta_L \right )   = h \frac{(d-2 \Delta_L)^2}{2 \Delta_L+1} \left ( \beta_R \dot{\beta}_L (\Delta_L+1) - \Delta_L \dot{\beta}_R \beta_L \right ).
\label{leftflux}
\end{equation}
Here we see that for the flux to be non-zero we explicitly need both $\alpha_L$ and $\beta_L$ to be non-zero, so ordinary boundary conditions, both standard and alternate, would indeed give a vanishing flux. In the final expression we chose to express the flux solely in terms of the $\beta$'s for future comparison. By energy conservation we have to find an equal opposite flux on the right hand side\footnote{Note that with our definition of the $z$ coordinate running from 0 to infinity on both sides, our normal vector points away from the boundary in both AdS spaces, so the left flux points to the left and the right flux points to the right. Hence the statement that energy conservation requires equal opposite fluxes.}.

\vskip10pt
\noindent{\bf Standard Quantization (Right):}
On the right side we run into an extra complication. Since the transparent boundary conditions no longer set $\alpha_R=0$ with the standard $k_R=0$ term we now have a divergent contribution already to the standard term:
\begin{equation}
\mathcal{F}^s_{R,\text{div}} =  (d - \Delta_R) \alpha_R \dot{\alpha}_R z^{d - 2\Delta_R}
= h^2 (d - \Delta_R) \beta_L \dot{\beta}_L z^{d - 2\Delta_R}.
\end{equation}
What we learn from this is that we need to add a non-vanishing improvement term on the right as well. Maybe not surprisngly, we need to add exactly the {\it same} coefficient
\begin{equation}
 k_R=k_L=- \frac{\Delta_L}{2 \Delta_L +1}.
\end{equation}
to keep the flux on the right side finite as well. With this the full flux once again reads
\begin{equation}
\mathcal{F}_R = \mathcal{F}_R^s + k_R \mathcal{F}^i_R.
\end{equation}
Note that for the standard boundary condition $\alpha_R=0$ the improvement term only receives a contribution proportional to $z^{2 \Delta_R -d}$ which vanishes as $\Delta_R > d/2$, which is consistent with the fact that usually the improvement term is only included for alternate boundary conditions. For the transparent boundary conditions it is however needed and does the job of cancelling the new divergence just as it did in the alternate case:
\begin{equation}
\mathcal{F}^{\text{div}}_R = \mathcal{F}^{\text{div},s}_R + \mathcal{F}^{\text{div},i}_R= z^{d-2\Delta_R} \,
\,  \alpha_R \dot{\alpha}_R \left [ (d- \Delta_R) + k_R (2 d - 2\Delta_R +1) \right ] = 0.
\end{equation}
With this it now it straightforward to calculate the finite flux:
\begin{equation}
\mathcal{F}_{R} = -  \mathcal{F}_{L} =- h \frac{(d-2 \Delta_L)^2}{2 \Delta_L+1} \left ( \beta_R \dot{\beta}_L (\Delta_L+1) - \Delta_L \dot{\beta}_R \beta_L \right ).
\end{equation}
As required by energy conservation, the fluxes are equal and opposite.

\subsection{Transmission Coefficients}

To calculate the transmission coefficient across the AdS boundary we need to solve the scalar wave equation for the scalar of mass $m$ and split it into an incoming, transmitted and reflected wave. The expression for the energy flux from the previous section will allow us to do the latter. As before,
we consider two copies of Poincaré patch AdS$_{d+1}$, labeled left (L) and right (R), each with a scalar field of mass $m$. The spacetime metric for each is
\begin{equation}
ds^2 = \frac{\ell^2}{z^2} \left( dz^2 - dt^2 + d\vec{x}^2 \right), \quad z > 0
\end{equation}
and we work in units where $\ell = 1$.

\subsubsection{Wave Equation and Separation of Variables}

The Klein-Gordon equation is:
\begin{equation}
\Box \phi - m^2 \phi = 0
\end{equation}
We take a separation of variables ansatz:
\begin{equation}
\phi(z, t, \vec{x}) = e^{-i\omega t + i \vec{k} \cdot \vec{x}} f(z)
\end{equation}
where it is implied that only the real part of  the right hand side corresponds to the physical field $\phi$ as is standard when solving linearized wave equations.
With this ansatz the radial equation becomes:
\begin{equation}
z^{d+1} \partial_z \left( z^{1-d} \partial_z f(z) \right) + \left( \omega^2 - \vec{k}^2 \right) z^2 f(z) - m^2 f(z) = 0
\end{equation}
Defining $q$ such that $\omega^2 = \vec{k}^2+q^2$ as we expect from a dispersion relation with parallel and perpendicular momentum for a wave impinging on $z=0$, and using $\nu = \sqrt{(d/2)^2 + m^2}$, the general solution is:
\begin{equation}
f(z) = z^{d/2} \left[ c_J J_\nu(qz) + c_Y Y_\nu(qz) \right]
\end{equation}
Here, $J_\nu$ and $Y_\nu$ are Bessel functions of the first and second kind.

For a scattering problem the standard Bessel functions are not a good basis. At large $z$ they behave like cosine and sine. A better linear combination is given by the Hankel functions
\begin{equation}
H^1_{\nu} = J_{\nu} + i Y_{\nu}, \qquad H^2_{\nu} = J_{\nu} - i Y_{\nu}
\end{equation}
which asymptotically, at large $z$, go as $e^{\pm i k z}$ and behave like a wave incoming from large $z$ ($H^2$) or outgoing to large $z$ ($H^1$). In terms of these we can equivalently write the most general solution to the wave equation as
\begin{equation}
f(z) = z^{d/2} \left[ c_1 H^1_\nu(qz) + c_2 H^2_\nu(qz) . \right]
\label{hsolution}
\end{equation}
While the interpretation of $c_1$ and $c_2$ as outgoing and incoming wave at large $z$ simply follows from the asymptotic form of these solutions, their interpretation at small $z$ is less apparent and we need to use the flux defined in the previous section to analyse this.

\subsubsection{Behavior at the Boundary}

As $z \to 0$
the solution \eqref{hsolution} has the standard form
\begin{equation}
\phi(z) \sim \alpha z^{d-\Delta} + \beta z^{\Delta}
\end{equation}
with $\Delta = d/2 + \nu$ for standard quantization and $\Delta = d/2 - \nu$ in alternate quantization. From the expansion of the Hankel functions we can read off $\alpha$ and $\beta$ on the Left (alternate) and Right (standard): 

\begin{align}
\alpha_R \mbox{ and } \beta_L &= i \left ( c_2^{L/R}-c_1^{L/R} \right ) \frac{2^{\nu} \Gamma[\nu]}{q^{\nu} \pi}\, e^{-i\omega t + i \vec{k} \cdot \vec{x}}, \label{asym1} \\
\beta_R \mbox{ and } \alpha_L &= q^{\nu}\frac{\left ( c_2^{L/R}+c_1^{L/R} \right ) \pi - i \left ( c_1^{L/R}-c_2^{L/R} \right ) \cos (\nu \pi) \Gamma[-\nu] \Gamma[1 +\nu]}{2^{\nu} \pi \Gamma[1+\nu]} \, e^{-i\omega t + i \vec{k} \cdot \vec{x}} \label{asym2}
\end{align}
To determine whether the linear combination in terms of the Hankel functions corresponds to incoming and reflected wave also at small $z$ we need to plug in $\alpha_L$ and $\beta_L$ into the formula \eqref{leftflux} for the left flux. We should keep in mind that only the real part of the complex expression we write for the field corresponds to the actual field value. Furthermore, like for calculations of the Poynting vector of an electromagnetic wave, the energy flux itself is time dependent and oscillates. But there is a non-trivial average flux which we can extract by averaging over a period
\begin{equation}
\langle \mathcal{F} \rangle = \frac{\omega}{2 \pi} \, \int_0^{2 \pi / \omega} dt \, \mathcal{F} (t) .
\end{equation}
Applying this to quantities of the form we are faced with,
\begin{equation}
 A_i = a_i e^{i \vec{k} \cdot \vec{x} - i \omega t}
\end{equation}
one finds that
\begin{eqnarray}
\langle A_1 \dot{A}_2 \rangle &=& \frac{1}{4}  \langle \left ( a_1 e^{i \vec{k} \cdot \vec{x} - i \omega t} +
a_1^* e^{-i \vec{k} \cdot \vec{x} +i \omega t} \right ) \left ( - i \omega a_2 e^{i \vec{k} \cdot \vec{x} - i \omega t} +
i \omega a_2^*  e^{-i \vec{k} \cdot \vec{x} + i \omega t} \right ) \rangle \\
&=& i \frac{\omega}{4} (a_1 a_2^* - a_2 a_1^*) = \frac{\omega}{2} \Im(a_1^* a_2)
\end{eqnarray}
where we used that $\langle e^{ \pm 2 i \omega t} \rangle =0$.

First let us apply this to the solution with $c_2=0$. The averaged flux in this case reads
\begin{equation}
\mathcal{F}_L = \frac{\omega}{2 \pi^2 } \frac{d-2\Delta_L}{2 \Delta_L+1} (2 \Delta_L+1 ) |c_1|^2 \frac{\Gamma[\nu]}{\Gamma[1+\nu]} = \frac{\omega}{ \pi^2 } |c_1|^2 \equiv  {\cal N}_L |c_1|^2
\end{equation}
where we collected all the positive prefactors into a single positive constant ${\cal N}_L$. The important part here is that since $|c_1|^2$ is positive, the imaginary part only gets contributions from taking the first term without the $i$ in $\alpha_L$. The second term in $\alpha_L$ gives a manifestly real contribution to $a_1^* a_2$ and so drops from the imaginary part. In contrast, setting $c_1=0$ we end up with
\begin{equation}
\mathcal{F}_L  = - {\cal N}_L |c_2|^2 .
\end{equation}
So the $H^1$ solution gives a positive flux, meaning flowing to large $z$, whereas the $H^2$ solution gives a negative flux, meaning flowing towards $z=0$. So reassuringly, $H^2$ and $H^1$ indeed retain their incoming and outgoing character near the boundary. Tagging the solution as incoming or outgoing at large $z$ is undesirable as it relies on the spacetime being empty AdS. We expect the transmission coefficient to be only a property of the boundary conditions. For any spacetime that is asymptotically AdS, near the boundary the solution is given in terms of the empty AdS Hankel functions and, reassuringly, we were able to confirm their incoming and outgoing character even near $z=0$ based on the associated energy fluxes.

Repeating the analysis on the right, $\alpha$ and $\beta$ switch their roles (which introduces an extra minus sign), but at the same time the flux on the right was minus the flux on the left, so we are left with the same conclusion: $H^2$ is incoming, whereas $H^1$ is outgoing also on the right.

\subsubsection{Transmission Across the Boundary}

We are now in a position to calculate the transmission coefficient across the boundary. When scattering from the left, we are looking at a solution with
\begin{equation}
c_2^R =0, \quad c_2^L=1, \quad c_1^L = r, \quad c_1^R =t
\end{equation}
With this we can evaluate the resulting near boundary coefficients in \eqref{asym1} and \eqref{asym2} and then solve the transparent boundary conditions \eqref{bcs} for $r$ and $t$:
\begin{eqnarray}
 t &=& h \frac{2 (d - 2 \Delta_L)}{(1+ (d-2\Delta_L)^2 h^2) (1 + i \cot(\pi (d/2 -  \Delta_L)))} \\
r &=& 1-   \frac{2 }{(1+ (d-2\Delta_L)^2 h^2) (1 + i \cot(\pi (d/2 -  \Delta_L)))}
\end{eqnarray}
and the corresponding transmission and reflection coefficients
\begin{eqnarray}
 |t|^2 &=& h^2 \frac{4 (d-2 \Delta_L)^2 \sin^2( \pi  (\frac{d}{2}-\Delta_L ))}{(1+ (d-2\Delta_L)^2 h^2)^2} \label{transandref} \\
|r|^2 &=& \cos^2( \pi  (\frac{d}{2}-\Delta_L )) -   \frac{(-1+(d-2 \Delta_L)^2 h^2)^2 \sin^2( \pi  (\frac{d}{2}-\Delta_L ))}{(1+ (d-2\Delta_L)^2 h^2)^2} 
\end{eqnarray}
Reassuringly $|r|^2+|t|^2=1$. It is also interesting to note that, like conformal interfaces in 2d \cite{Quella:2006de,Meineri:2019ycm}, the result does not depend on the properties of the wave, in particular it is independent of $q$ and $\omega$.

\begin{figure}
    \centering
    \includegraphics[width=0.5\linewidth]{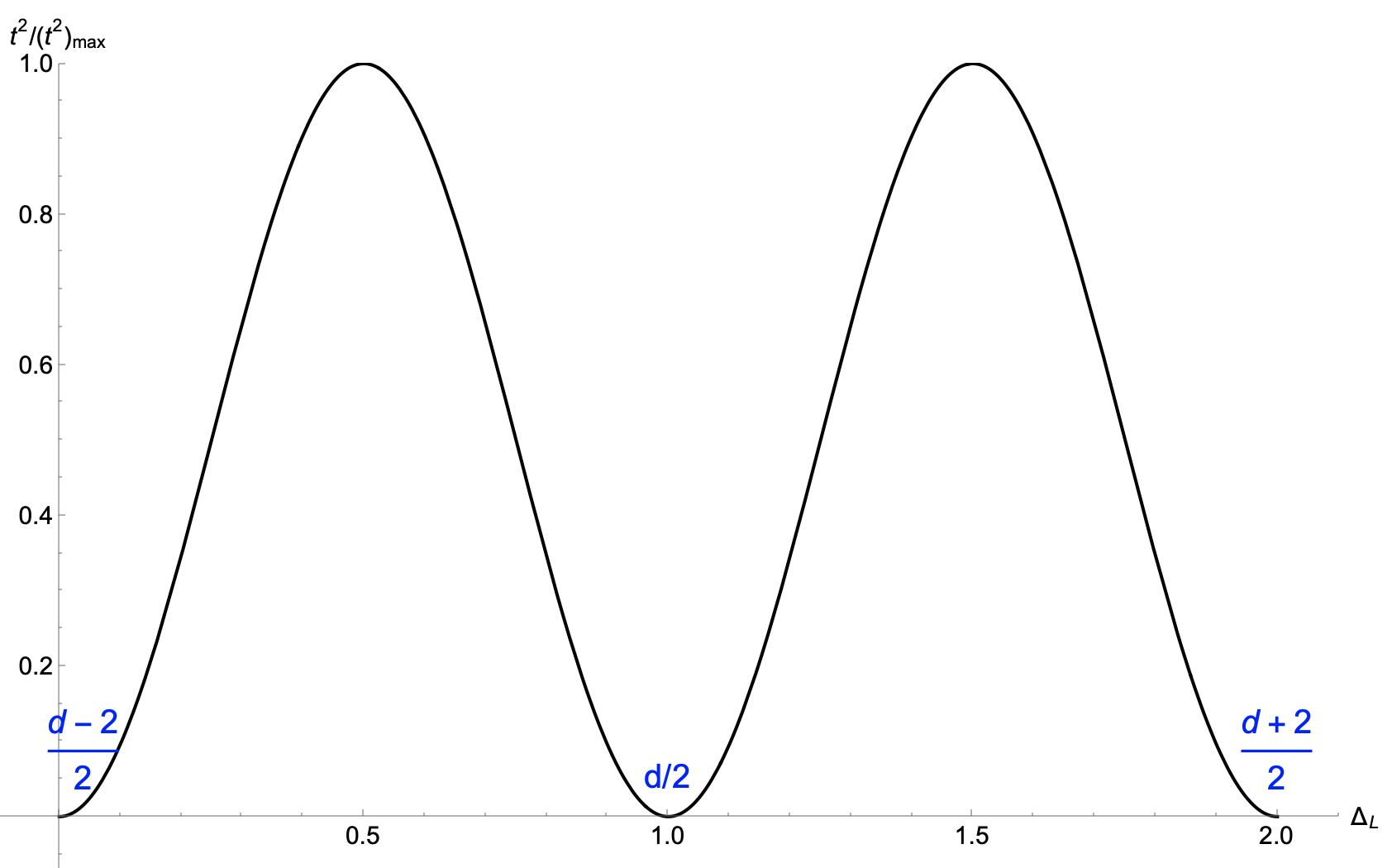}
    \caption{Transmission coefficient across the transparent boundary conditions.}
    \label{fig:trans-coeff}
\end{figure}

While the general expressions are quite cumbersome, let us consider a few special cases. 
First note, that the transmission coefficient peaks at
\begin{equation}
h^2_{max} = \frac{1}{(d-2 \Delta_L)^2}
\label{hmax}.
\end{equation}
At this value of $h$ we find that the transmission coefficient is given by
\begin{equation}
|T^2|_{h_{max}}= \sin^2( \pi  (\frac{d}{2}-\Delta_L )).
\end{equation}
which, as it needs to be, is always bounded above by 1 and saturates to 1 at $\Delta_L=(d-1)/2$, the conformally coupled scalar.
In contrast, linearizing in $h$ we find
\begin{equation}
|t|^2 = 4 h^2  (d-2\Delta_L)^2 \sin^2( \pi  (\frac{d}{2}-\Delta_L )), \qquad
|r|^2 = 1- 4 h^2  (d-2\Delta_L)^2 \sin^2( \pi  (\frac{d}{2}-\Delta_L )).
\end{equation}
This makes the results a little more transparent. The transmission of course vanishes if $h$ vanishes. See Figure \ref{fig:trans-coeff}. As a function of dimension, it vanishes at the edges of the allowed interval,
\begin{equation}
\Delta_L^{min} = \frac{d-2}{2}, \qquad \Delta_L^{max} = \frac{d}{2}
\end{equation}
and peaks once again in the middle, which is given by the conformally coupled scalar
\begin{equation}
\Delta_L^{conf} = \frac{d-1}{2}.
\end{equation}
The conformally coupled scalar is a very special case to consider. Since for the conformally coupled scalar we can conformally map AdS to Minkowski space, in this case we can check our result against a conformal interface in flat space. Let us record the result for the conformally coupled scalar and we will compare to flat space in the next subsection:
\begin{equation}
|t|^2_{conf} = \frac{4}{ (h+h^{-1})^2}, \qquad |r|_{conf}^2 = \frac{ (h - h^{-1})^2}{(h+h^{-1})^2} . \label{conformal}
\end{equation}

\subsubsection{Comparison to Free Scalar in Flat Space}

We consider a plane wave of a free scalar field incident at an angle on a conformal planar interface separating two regions. The field and its normal derivative transform via an SL(2,$\mathbb{R}$) matrix \cite{Kim:2025tvu}
\begin{equation}
\begin{pmatrix}
 \phi \\
\partial_x\phi
\end{pmatrix}_{x=0^+}
=
\begin{pmatrix}
\lambda & 0 \\
0 & \frac{1}{\lambda}
\end{pmatrix}
\begin{pmatrix}
 \phi \\
\partial_x \phi
\end{pmatrix}_{x=0^-}
\end{equation}
As shown in \cite{Kim:2025tvu} this is the general form for a conformal defect in  a more than $d=2$ dimensional ambient space. For a general 2 x 2 matrix with entries $a$, $b$, $c$, $d$, energy conservation demands $ad-bc=1$. Scale invariance further restricts us to the form we use with $b=c=0$. Conformal invariance then comes for free. In $d=2$ the situation is different as was analyzed in \cite{Bachas:2001vj}. There one can impose matching conditions which employ the time derivative of the field rather than the field value, allowing one a rotation angle rather than the single real scaling parameter $\lambda$. Such a boundary condition in general $d$ is not consistent with the Lorentz invariance along the slice.

The incoming, reflected, and transmitted waves take the form:
\begin{equation}
\phi_L = e^{-i \omega t} e^{i \vec{k}_\parallel \cdot \vec{x}_\parallel} \left( e^{i k_\perp x} + r\, e^{-i k_\perp x} \right), \quad
\phi_R = t\,e^{-i \omega t}  e^{i \vec{k}_\parallel \cdot \vec{x}_\parallel}  e^{i k_\perp x}
\end{equation}
with
 \begin{equation}
 \omega^2 = k_\parallel^2 + k_{\perp}^2 .
 \end{equation}
All fields and their derivatives have an overall $e^{-i \omega t} e^{i \vec{k}_\parallel \cdot \vec{x}_\parallel}$ dependence, so if they obey the boundary conditions for one point on the interface they obey them for all. So without loss of generality we can set $t=\vec{x}_\parallel=0$ for the purpose of imposing the boundary conditions and will do so from now on. Evaluating  the field and its normal derivative at \( x = 0 \), we have:
 \begin{equation}
\partial_t \phi_L = - i \omega ( 1 + r), \quad \partial_x \phi_L = i k_\perp (1 - r)
 \end{equation}
 \begin{equation}
\partial_t \phi_R = - i \omega t, \quad \partial_x \phi_R = i k_\perp t
 \end{equation}

Matching these via the SL(2,$\mathbb{R}$) matrix yields a system of equations that can be solved for \( r \) and \( t \).

The resulting reflection and transmission amplitudes are:
 \begin{equation}
r = \frac{\lambda - \lambda^{-1}}{\lambda + \lambda^{-1}}, \qquad
t = \frac{2}{\lambda+\lambda^{-1}} \qquad r^2 + t^2 =1.
 \end{equation}
Remarkably, just as in the case of 2d CFTs, these transmission amplitudes are independent of the properties of the incoming wave, in particular $k_\perp$ and $k_\parallel$.
The resulting $|r|^2$ and $|t|^2$ are in perfect agreement with the holographic result \eqref{conformal} for the conformally coupled scalar if we identify $\lambda$ with $h$.
Repeating the scattering analysis for a wave incoming from the right one finds the same value for $t$ and hence the same transmission and reflection coefficients.

\section{A Novel Strong/Weak Coupling Duality for Holographic Open Systems}

In our general prescription of open holography from section \ref{open} it was important that

\begin{itemize}
\item The coupling from system to bath was via a marginal coupling, $\Delta_L = d- \Delta_R$.
\item As a consequence, both $\Delta_{L/R}$ have to live in the window
\begin{equation}
\frac{d-2}{2} \leq \Delta_{L/R} \leq \frac{d+2}{2}
\end{equation}
where two quantizations are possible.
\item Hence one of the sides has to employ standard quantization, $\Delta> d/2$, whereas the other side has alternate quantization, $\Delta < d/2$
\end{itemize}

Let us denote the partition function of a theory obeying the 3 requirements above by
$Z[{\Delta,d-\Delta,h}]$.
The first and second entry are left and right dimensions. They automatically implement the structure that they require opposite quantizations. We claim that the system exhibits a strong/weak coupling duality: 
\begin{equation}
Z[\Delta,d-\Delta,h] = Z[d-\Delta,\Delta,\frac{1}{h} \frac{1}{(2 \Delta-d)^2}] . \label{duality}
\end{equation}
That is, we can map the theory at strong coupling to the theory at weak coupling with the role of standard and alternate quantization reversed. 

The duality is easy to establish on the gravity side. Given that it is completely universal in the sense that none of the details of the CFT matter, we suspect it to be true in any large $N$ setting, irrespective of whether it has a gravity dual.

To proceed, we first rewrite the boundary conditions \eqref{bcs} as
\begin{equation}
\alpha_L =  -h (2 \Delta_L-d) \beta_R, \quad \alpha_R =  h (2 \Delta_L -d) \beta_L \label{bcsnew}
\end{equation}
where we simply used $\Delta_R = d-\Delta_L$ in \eqref{bcs}. In terms of the rescaled coupling 
\begin{equation}
\label{rescaled}
g= (2 \Delta_L -d) h
\end{equation}
this simply becomes
\begin{equation}
\alpha_L[\Delta_L] = -g \beta_R[d-\Delta_L] , \quad \alpha_R[d-\Delta_L] =  g \beta_L[\Delta_L] . \label{bcsnewnew}
\end{equation}
This time we also explicitly displayed the dimension we use on the left and right when defining $\alpha$ and $\beta$.

Last but not least we use the fact that changing standard with alternate quantization simply exchanges the role of $\alpha$ and $\beta$, that is
\begin{equation}
\label{identity}
\alpha[d-\Delta] = \beta[\Delta], \quad \beta[d-\Delta] = \alpha[\Delta] .
\end{equation}
With this it is easy to see what happens to our boundary conditions \eqref{bcsnewnew} under the proposed duality transformation
\begin{equation}
\label{asswap}
\Delta_L \rightarrow  d - \Delta_L, \quad \Delta_R = d- \Delta_L \rightarrow d-\Delta_R = \Delta_L, \quad
h \rightarrow \frac{1}{h} \frac{1}{(2 \Delta_L-d)^2}, \quad
g \rightarrow -\frac{1}{g}.
\end{equation}
The transformation of $g$ follows from the transformation of $h$ and $\Delta_L$:
\begin{equation}
g_{new} = (2\Delta_{L,new} -d) h_{new} = - (2 \Delta_L -d) \frac{1}{h} \frac{1}{(2 \Delta_L-d)^2} = - \frac{1} {h (2 \Delta_L -d)} = - \frac{1}{g}.
\end{equation}
The boundary conditions \eqref{bcsnewnew} after the duality transformation become
\begin{equation}
\alpha_L(d-\Delta_L) = \frac{1}{g} \beta_R(\Delta_L) , \quad \alpha_R(\Delta_L) =  -\frac{1}{g} \beta_L(d-\Delta_L) .\label{bcsswapped}
\end{equation}
which after using the identity \eqref{identity} read 
\begin{equation}
\beta_L(\Delta_L) = \frac{1}{g} \alpha_R(d-\Delta_L) , \quad \beta_R(d-\Delta_L) =  -\frac{1}{g} \alpha_L(\Delta_L) .\label{bcsswapped2}
\end{equation}
which are clearly identical to the original boundary conditions \eqref{bcsnewnew}. This establishes the duality \eqref{duality}.

As a sanity check, it is illuminating to look at the case of very small and very large $h$. In the case that $h \rightarrow 0$ the transparent boundary conditions \eqref{bcs} reduce to $\alpha_L = \alpha_R =0$, two decoupled scalars with their chosen quantizations. In contrast, very large $h$ instead sets $\beta_L=\beta_R=0$. These are also familiar boundary conditions, but they reverse the quantization choice. Our duality calculation demonstrates that this duality in fact holds for the entire range of $h$.

Another nice consistency check is that the transmission coefficient \eqref{transandref} is in fact invariant under the duality relation, and the maximal value \eqref{hmax} for $h$ is exactly the self-dual coupling.

\section{Review of Quasinormal Modes Before Coupling to Bath}

In order to study how an originally closed system turns into a dissipative system upon coupling to the bath, and to quantify the resulting dissipative time scales, we want to determine the (quasi) normal modes of the system, that is the eigen frequencies of normalizable excitations living on the geometry. Without the coupling to the bath, a CFT at zero temperature exhibits unitary time evolution and all eigenfrequencies are real. The wave operator in the dual bulk is Hermition and so in fact has standard normal modes. At finite temperature the bulk exhibits a black hole, the normal modes turn into quasi normal modes and the corresponding imaginary part of the eigenfrequencies quantifies the exponential decay of excitations in a thermal system. Our aim is to quantify how the normal modes of the zero temperature system turn into quasi-normal modes upon coupling to a finite temperature bath. In particular, the least negative imaginary part will give us the leading dissipative time scale. But before we can present this analysis, we need to first review what is known about normal modes and quasi normal modes in AdS and AdS black hole backgrounds.

\subsection{Normal Modes for Empty AdS}
We start from the metric of empty AdS of d+1 dimensions:
\begin{equation}
    ds^2 = -f(r)dt^2+\frac{1}{f(r)}dr^2+r^2 d\Omega_{d-1}
\end{equation}
The Klein Gordon equation looks like: 
\begin{equation}
    (\Box-m^2)\Phi=0 \qquad \text{where}\quad \Box =\frac{1}{\sqrt{-g}} \partial_\mu (\sqrt{-g} g^{\mu\nu}\partial_\nu \Phi) 
\end{equation}
A separation of variables ansatz for $\Phi$: $\Phi = R(r)e^{-i \omega t }Y(\phi)$
yields
\begin{equation}
\begin{split}
    (\Box-m^2)\Phi&=r^{1-d} \partial_\mu (r^{d-1}g^{\mu\nu}\partial_\nu \Phi)\\ &= \frac{\omega^2}{f(r)}R Y e^{i \omega t } + r^{1-d} \partial_r (r^{d-1} f(r) R'(r) Y e^{-i \omega t})+\frac{1}{r^2}(d-1)(\partial_\phi\partial_\phi \Phi)-m^2 \Phi = 0
\end{split}
\end{equation}
The third term in the equation above can be rewritten in an eigen equation form.
\begin{equation}
    \frac{1}{r^2}(d-1)(\partial_\phi\partial_\phi \Phi) = \frac{1}{r^2}(\Box_{s^{d-1}} Y) R e^{-i \omega t} = -\frac{1}{r^2} l(l+d-2)Y R e^{-i \omega t} 
\end{equation} 
After simplifying, the final equation we have is:
\begin{equation}
    \frac{\omega^2}{f(r)}R(r) + r^{1-d} \partial_r (r^{d-1} f(r) R'(r))-\frac{1}{r^2}l(l+d-2) R(r)-\Delta(\Delta-d) R(r) = 0
\end{equation}
with
\begin{equation}
    \Delta_\pm = \frac{1}{2}(d\pm\sqrt{d^2+4 m^2})
\end{equation}
being the two possible values for $\Delta$ in standard and alternate quantization respectively.
This equation will have two solutions of hypergeometric form. One of them is singular near the origin at $r=0$ which forces us to set it  equal to 0. The other we expand near the boundary. From the expansion we get the $\alpha$ and $\beta$ as:
\begin{align}
\alpha &= \frac{\Gamma \left(\Delta -\frac{d}{2}\right) \Gamma \left(\frac{d}{2}+l\right)}{\Gamma \left(\frac{1}{2} (l-w+\Delta )\right) \Gamma \left(\frac{1}{2} (l+w+\Delta )\right)} \\
\beta &= \frac{\Gamma \left(\frac{d}{2}-\Delta \right) \Gamma \left(\frac{d}{2}+l\right)}{\Gamma \left(\frac{1}{2} (d+l-w-\Delta )\right) \Gamma \left(\frac{1}{2} (d+l+w-\Delta )\right)}
\end{align}
If we take $\alpha$ to be the non-normalizable mode, then the modes coming from the poles of this coefficient will have the form \cite{Aharony:1999ti}:
\begin{equation}
    \omega/L = \pm (l+\Delta+2n).
\end{equation}

\subsection{Quasinormal Modes of the Spinning BTZ Black Hole}
While the normal modes of AdS we reviewed in the previous subsection can be found analytically in any $d$, the general AdS black hole does not allow closed form expressions for the quasi normal mode. Analytic answers can however be found \cite{Birmingham_2001} for the special case of the BTZ black hole in $d=2$.
Let's start with the metric in a BTZ background
\begin{equation}
    ds^2 = -(-M +\frac{r^2}{l^2}+\frac{J^2}{4 r^2})dt^2+(-M +\frac{r^2}{l^2}+\frac{J^2}{4 r^2})^{-1} dr^2 +r^2(d\phi-\frac{J}{2 r^2}dt^2)^2
\end{equation}
with $M = \frac{r_+^2+r_-^2}{l^2},J=\frac{2 r_+r_-}{l}$
the scalar wave equation is 
\begin{equation}
    (\nabla^2-\frac{\mu^2}{l^2})\Phi=0
\end{equation}
using the ansatz $\Phi=R(r)e^{-I \omega t} e^{i m \phi}$. Now the equation look like
\begin{equation}
\begin{split}
(\Box-\mu^2)\Phi &= \frac{\frac{\partial }{\partial r}\left(r \frac{\partial \Phi }{\partial r} \left(\frac{J^2}{4 r^2}+\frac{r^2}{l^2}-M\right)\right)}{r}+\frac{r^2 \frac{\partial }{\partial t}\frac{\partial \Phi }{\partial t}}{-\frac{J^2}{4}-\frac{r^4}{l^2}+M r^2}+\frac{J \frac{\partial }{\partial \phi }\frac{\partial \Phi }{\partial t}}{2 \left(-\frac{J^2}{4}-\frac{r^4}{l^2}+M r^2\right)}\\ 
&+\frac{J \frac{\partial }{\partial t}\frac{\partial \Phi }{\partial \phi }}{2 \left(-\frac{J^2}{4}-\frac{r^4}{l^2}+M r^2\right)}+\frac{\left(M-\frac{r^2}{l^2}\right) \frac{\partial }{\partial \phi }\frac{\partial \Phi }{\partial \phi }}{-\frac{J^2}{4}-\frac{r^4}{l^2}+M r^2}-\frac{\mu^2  \Phi }{l^2}=0
\end{split}
\end{equation}
With the change of variables\footnote{This is not the same $z$ coordinate we used before, in particular the boundary is now at $z=1$. But since it is standard in the mathematical physics literature to call the argument of Hypergeometric functions $z$ we briefly will give $z$ this new meaning.} $z=\frac{r^2-r_+^2}{r^2-r_-^2}$ the radial equation takes the form 
\begin{equation}
    z(1-z)\frac{d^2R}{dz^2}+(1-z)\frac{dR}{dz}+(\frac{A}{z}+B+\frac{C}{1-z})R=0
\end{equation}
With 
\begin{equation}
\begin{split}
A &= \frac{l^4 r_+^2 \omega ^2-2 l^3 m r_- r_+ \omega +l^2 m^2 r_-^2}{4 (r_--r_+)^2 (r_-+r_+)^2} \\
B&= -\frac{l^2 (l r_- \omega -m r_+)^2}{4 \left(r_-^2-r_+^2\right)^2}\\
C&=\frac{-\mu^2}{4}
\end{split}
\end{equation}
To put the hypergeometric function in a canonical form, we define $R(z) = z^\alpha(1-z)^\beta F(z)$. The radial equation now looks
\begin{equation}
    z(1-z)\frac{d^2F}{dz^2}+[c-(1+a+b)z]\frac{dF}{dz}- ab F=0
\end{equation}
where 
\begin{equation}
\begin{split}
    c &= 2\alpha +1\\
    a+b &= 2\alpha +2\beta \\
    a\,b&=(\alpha+\beta)^2-B \\
    \implies a &=\alpha +\beta -\sqrt{B}\\
    b&=\alpha +\beta +\sqrt{B}
\end{split}
\end{equation}
for which $\alpha = - i \sqrt{A}$ and $\beta=\frac{1}{2} (1- \sqrt{1+\mu^2})$. without loss of generality, we choose $\alpha^2 = -A$ and $\beta=\frac{1}{2} (1\pm \sqrt{1+\mu^2})$. The purely ingoing solution to the hypergeometric function picks one of the two independent solutions:
\begin{equation}
    R(z) = z^\alpha(1-z)^\beta {}_2F_1(a,b,c,z)
\end{equation}
To extract the near boundary coefficients $\alpha$ and $\beta$ we will expand $R(z)$ near the boundary $z=1$ to obtain the coefficients matching the expected field profile: \footnote{Due to the change of coordinates made above where $z=\frac{r^2-r_+^2}{r^2-r_-^2}$, the powers of the expansion becomes $ \alpha \, (\sqrt{1-z})^{2-\Delta}+ \beta \, (\sqrt{1-z})^\Delta$. We could always define $x=\sqrt{1-z}$ where now $x\to 0$ as $z\to1$ near the boundary.} 

\begin{equation}
\begin{split}
 \alpha&=-\frac{\pi \, \csc (\pi  \Delta) \,\Gamma \left(\frac{i l (l r_+ \omega -m r_-)}{r_-^2-r_+^2}+1\right)}{\Gamma (2-\Delta)\, \Gamma \left(\frac{1}{2} \left(\Delta+\frac{i l (l \omega -m)}{r_- -r_+}\right)\right)\, \Gamma \left(\frac{1}{2} \left(\Delta-\frac{i l (m+l \omega )}{r_-+r_+}\right)\right)}\\
\beta&=\frac{\pi \, \csc (\pi  \Delta) \, \Gamma \left(\frac{i l (l r_+ \omega -m r_-)}{r_-^2-r_+^2}+1\right)}{\Gamma (\Delta) \, \Gamma \left(\frac{1}{2} \left(-\Delta+\frac{i l (l \omega -m)}{r_--r_+}+2\right)\right) \, \Gamma \left(\frac{1}{2} \left(-\Delta-\frac{i l (m+l \omega )}{r_-+r_+}+2\right)\right)}.
\label{eq:coeffsmassivespinBTZ}
\end{split}
\end{equation}
Note that we have made a choice during the expansion, that is $\sqrt{\mu^2+1}=+(\Delta_+-1)$ which enforces the standard quantization. To ensure no divergences near the boundary, one conventionally sets the non normalizable mode to 0 which can be done through the poles of $\alpha$ \cite{Birmingham_2001}:
\begin{equation}
    \begin{split}
        \omega_1 &=\frac{m}{l}-2 i \Big( \frac{r_+-r_-}{l^2}\Big)\Big(n+\frac{1}{2}+\frac{1}{2}\sqrt{1+\mu^2}\Big)\\
        \omega_2 & =-\frac{m}{l}-2 i \Big( \frac{r_++r_-}{l^2}\Big)\Big(n+\frac{1}{2}+\frac{1}{2}\sqrt{1+\mu^2}\Big)
    \end{split}
\end{equation}
Setting $\alpha = 0$ corresponds to the reflecting boundary conditions of AdS. We will analyze the case of transparent boundary conditions in the next section.

\subsection{Quasinormal Modes of the Schwarzschild BTZ Black Hole}
To get the leading and the subleading coefficient of the scalar field profile near the boundary of AdS with a standard Schwarzschild BTZ from the spinning BTZ \eqref{eq:coeffsmassivespinBTZ} we can simply set $J=0, r_- = 0$ and $r_+ = l \sqrt{M}$ where $l$ here is the AdS length (we set it =1):
\begin{equation}
\begin{split}
    \alpha &= -\frac{\pi \, \csc (\pi  \Delta) \, \Gamma \left(1-\frac{i l \omega }{\sqrt{M}}\right)}{\Gamma (2-\Delta) \, \Gamma \left(\frac{1}{2} \left(\Delta-\frac{i (l \omega -m)}{\sqrt{M}}\right)\right) \, \Gamma \left(\frac{1}{2} \left(\Delta-\frac{i (m+l \omega )}{\sqrt{M}}\right)\right)}\\
    \beta &= \frac{\pi \, \csc (\pi  \Delta)\, \Gamma \left(1-\frac{i l \omega }{\sqrt{M}}\right)}{\Gamma (\Delta)\, \Gamma \left(\frac{1}{2} \left(-\Delta-\frac{i (l \omega -m)}{\sqrt{M}}+2\right)\right)\, \Gamma \left(\frac{1}{2} \left(-\Delta-\frac{i (m+l \omega )}{\sqrt{M}}+2\right)\right)}
\end{split}
\end{equation}
Using the above coefficients, the quasi normal modes for massive scalar perturbation on a Schwarzschild black hole background become:
\begin{equation}
    \omega = \pm m - 2 i \sqrt{M} (n+\frac{\Delta}{2})
\end{equation}
which matches the quasi normal frequencies in \cite{Cardoso:2001hn} for massless scalar perturbations, i.e. $\Delta = 2$.  

\section{Quasinormal Modes for the Open System}

\subsection{General Setting}

Our goal is to see what happens to the quasinormal modes in the open system of the CFT coupled to the bath. That is, we want to impose the boundary conditions \eqref{bcs} and repeat the analysis of the previous section. 

One note of caution: in the previous sections we obtained the coefficients $\alpha$ and $\beta$ for a given value of $\Delta$ and bulk temperature $T$ (and even angular momentum $J$, but let us ignore this for now, it is easy to incorporate this into this analysis). Since we solved linear equations, there is always the ambiguity that one could rescale $\alpha$ and $\beta$ by the same overall constant $C$ and still get a solution. When one imposes boundary conditions on one side only, say $\alpha_L=0$, the overall scaling of $\alpha_L$ is irrelevant: any frequency that ensures $\alpha_L=0$ also sets $C \alpha_L=0$. But now that we want to work with boundary conditions \eqref{bcs} that equate left and right quantities, a rescaling
\begin{equation}
\alpha_R \rightarrow C \alpha_R, \quad  \beta_R \rightarrow C \beta_R
\end{equation}
does not change the fact that we solved the equations of motion on both sides but clearly matters when solving the boundary conditions. Only a simultaneous rescaling of both left and right solutions drops out.

To account for this let us introduce some notation. We denote by $\alpha(\Delta,T)$ and $\beta(\Delta,T)$ the standard solutions for the quasinormal modes in the corresponding black holes as reviewed in the previous subsection. We are looking for solutions of the form
\begin{eqnarray}
\alpha_L &=& \alpha(T_L,d-\Delta_R), \quad \beta_L = \beta(T_L, d-\Delta_R), \\ \quad \alpha_R &=& C \alpha(T_R,\Delta_R), \quad 
\beta_R = C \beta (T_R,\Delta_R).
\end{eqnarray}
That is we are expressing {\it both} sides in terms of the set of same basic quasinormal modes with the respective temperature, just with an overall rescaling on the right side. With this the boundary conditions \eqref{bcs} as \begin{eqnarray}
\alpha_L(T_L,d-\Delta_R) &=&  C\,  h (2 \Delta_R-d) \beta_R(T_R,\Delta_R), \quad \\ C \, \alpha_R(T_R,\Delta_R)  &=&  -h (2 \Delta_R -d) \beta_L(T_L,d-\Delta_R) \label{freebcs}.
\end{eqnarray}

The constant $C$ in \eqref{freebcs} is absolutely essential to get a consistent system of equations while using the results from the previous subsection. We have two equations to solve for the two unknowns: $C$ and the mode frequency $\omega$. The physical meaning of $C$ is that it tells us where the wave function of the mode is localized. A small $C$ indicates that the ``true" $\alpha_R$, which in \eqref{freebcs} is $C$ times the ``standard" $\alpha_R$ is very small: the modefunction is mostly living in the left AdS, but has a small tail in the right AdS whose amplitude is suppressed by $C$. Similarly, a very large $C$ tells us we find a mode that is mostly living in the right AdS with a small tail in the left.

\subsection{Equal Temperatures}

As a first scenario we can analytically solve, let us consider the case where both the left and right black hole have the same temperature (potentially 0). Without loss of generality let us continue to consider the case where $\Delta_R \geq d/2$, that is the right is standard quantized. In this case the boundary condition \eqref{freebcs} reads  \begin{equation}
\alpha(T,d-\Delta_R) =   C\,  h (2 \Delta_R-d) \beta(T,\Delta_R), \quad C \alpha(T,\Delta_R) = -h (2 \Delta_R -d) \beta(T,d-\Delta_R) \label{freebcs2}
\end{equation}
which using the identity \eqref{identity} between the expressions with dimension $\Delta_R$ and $d-\Delta_R$ simply becomes  \begin{equation}
\alpha(T,d-\Delta_R) [1 - C h (2 \Delta_R-d) ] =0 , \quad \alpha(T,\Delta_R) [C+ h (2 \Delta_R-d) ] =0 
\label{freebcs3}.
\end{equation}
These equations can easily be solved. We see there are two solutions
\begin{equation}
    \alpha(T,d-\Delta_R)=0, \qquad C = -h ( 2\Delta_R-d)
\end{equation}
and 
\begin{equation}
    \alpha(T,\Delta_R)=0, \qquad C =  \frac{1}{ h (2\Delta_R-d)}
\end{equation}
The first yields exactly the frequencies of an alternate quantized scalar with dimension $d-\Delta_R$ with a value of $C$ that is small for small $h$. That is, all the left quasinormal modes survive with unchanged frequencies. All that happens is their modefunction has some leakage to the right which is small for small transmission. The second in turn yields exactly the frequencies of a standard quantized scalar with dimension $\Delta_R$ and a large $C$ at small $h$. This time the right quasinormal frequencies survive unchanged. Since this time $C$ is large for small $h$ their modefunctions are mostly localized on the right but have some small leakage to the left.

To see genuinely new frequencies we have to work with unequal temperatures. We will turn to that case in the next section.

\section{Numerical Results}

\begin{figure}
    \centering
    \includegraphics[width=0.8\linewidth]{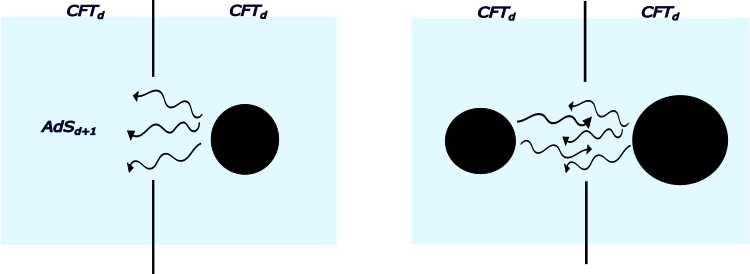}
    \caption{On the left a black hole coupled to empty AdS and on the right we couple two black holes.}
    \label{fig:placeholder}
\end{figure}

To get interesting quasi-normal modes we just saw we need to have un-equal size black holes on the two sides. In this case we can no longer give closed form solutions and rely on numerical analysis. We consider two cases: first we consider the special scenario with one zero temperature CFT coupled to a finite temperature bath, and later we generalize to the case of two non-zero but unequal temperatures. The two options are illustrated in Fig. \ref{fig:placeholder}.

\subsection{A Schwarzschild Black Hole Coupled to Empty AdS}
For completeness, we list the coefficients involved in every computation. For a Schwarzschild black hole with standard quantization:
\begin{equation}
    \begin{split}
        \alpha_R &=-\frac{\pi  \csc (\pi  \Delta ) \Gamma \left(1-\frac{i \omega }{\sqrt{M}}\right)}{\Gamma (2-\Delta ) \Gamma \left(\frac{1}{2} \left(\Delta -\frac{i (\omega -m)}{\sqrt{M}}\right)\right) \Gamma \left(\frac{1}{2} \left(\Delta -\frac{i (m+\omega )}{\sqrt{M}}\right)\right)}\\
        \beta_R &=\frac{\pi  \csc (\pi  \Delta ) \Gamma \left(1-\frac{i \omega }{\sqrt{M}}\right)}{\Gamma (\Delta ) \Gamma \left(\frac{1}{2} \left(-\Delta -\frac{i (\omega -m)}{\sqrt{M}}+2\right)\right) \Gamma \left(\frac{1}{2} \left(-\Delta -\frac{i (m+\omega )}{\sqrt{M}}+2\right)\right)}.
    \end{split}
    \label{eq:bhcoeff}
\end{equation}

For an empty AdS with alternate quantization:
\begin{equation}
\begin{split}
 \alpha_L &= \frac{\Gamma (1-\Delta ) \Gamma (l+1)}{\Gamma \left(\frac{1}{2} (l-\omega-\Delta +2)\right) \Gamma \left(\frac{1}{2} (l+\omega-\Delta +2)\right)} \\
 \beta_L &= \frac{\Gamma (\Delta -1) \Gamma (l+1)}{\Gamma \left(\frac{1}{2} (l-\omega+\Delta )\right) \Gamma \left(\frac{1}{2} (l+\omega+\Delta )\right)}.
\end{split}
\label{eq:emptyAdsCoeff}
\end{equation}
\begin{figure}
    \centering
    \includegraphics[width=0.45\linewidth]{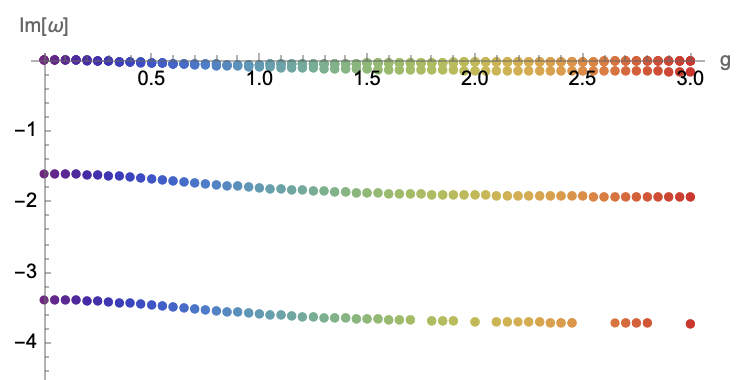}
    \includegraphics[width=.45\linewidth]{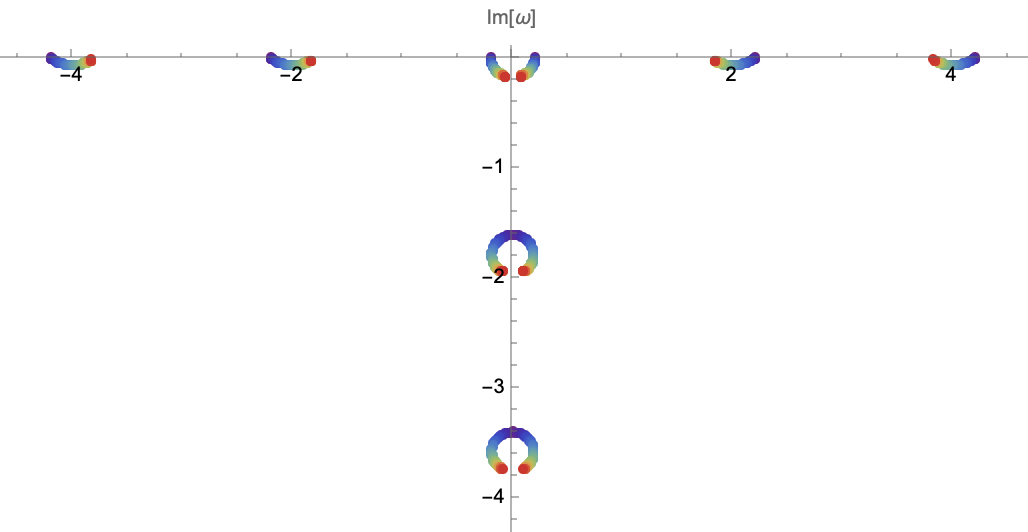}
    \caption{The parameters for this plot are: $\Delta= 1.8$, $M=.8$. The colors correspond to 61 different values of $g$: the purple correspond to $g=0$ the red correspond to $g=3$ and in between we increase the value of g in increments of $.05$. The duality between $g$ and $1/g$ is manifest: the bluer the dots are, the smaller g is and the closer we get to the modes associated with the $\alpha =0$ for standard quantization for the black hole and $\beta=0$ for standard quantization for empty AdS. The larger g gets, the redder the dots are and the closer we get to the modes associated with $\beta=0$ for standard quantization for the BH and $\alpha=0$ for standard quantization for empty AdS.}   \label{fig:1bhmodesdiffg}
\end{figure}
Eq. \eqref{bcs} for a black hole coupled to empty AdS is:
\begin{equation}
\alpha_R\left(\omega,\Delta,m,M\right) \alpha_L\left(\omega,\Delta,m\right)-h^2 (2\Delta_L-d)(2\Delta_R-d) \beta_R\left(\omega,\Delta,m,M\right) \,\beta_L\left(\omega,\Delta,m\right)=0\label{eq:emptyandbh}
\end{equation}
where $\alpha_R \,$and $  \beta_R$ have 4 parameters: $\omega, \Delta =1.8, m=0, M=.8$ respectively while  $\alpha_L \,\text{and}\, \beta_L$ have 3 parameters: $\omega, \Delta=1.8, l=0$ where $l$ and $m$ are the quantum number for empty AdS and for the BH, respectively. We set them to be the same value on both sides. Since we take the black hole to be on the right with standard quantization, the empty AdS will have alternate quantization which (for convenience) means we switch the coefficients we obtain for the field profile with standard quantization for empty AdS.

To explore how quasi normal modes of the coupled system respond to each of its parameters, we hold two of them fixed (we have already set the quantum number $m=0$) and we vary over the other. In Fig.  \ref{fig:1bhmodesdiffg}, we hold $M$ and $\Delta$ fixed and we vary the strength of the coupling between the black hole and the empty AdS. Interestingly, the loops in this figure is a clear manifestation of the strong/weak coupling duality: the quasi normal modes of the black hole start from the imaginary axis and make a full loop back to the imaginary axis again but with the new quantization. In other words, increasing the coupling drives the system from standard quantization on the right and alternate on the left to alternate on the right and standard on the left.
\begin{figure}
    \centering
    \includegraphics[width=0.5\linewidth]{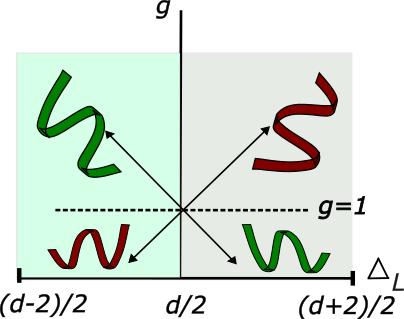}
    \caption{Phase diagram for the strong/week duality. We show that we don't need to cover the whole range of what could $\Delta_L$ be as long as we are running the coupling over large values above 1. The modes of the left side of the diagram covers the whole range of possible modes. The quasinormal modes of the bottom left side is the same as those of the top right and those of the bottom right is the same as those of the top left.}
    \label{fig:strong-week-duality}
\end{figure}

\begin{figure}
    \centering
    \includegraphics[width=0.3\linewidth]{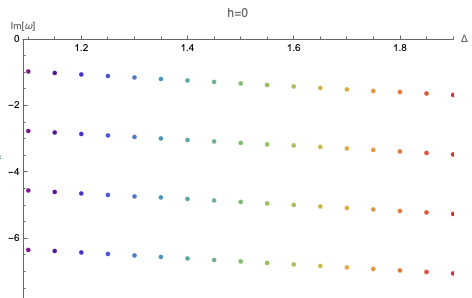}
    \includegraphics[width=0.345\linewidth]{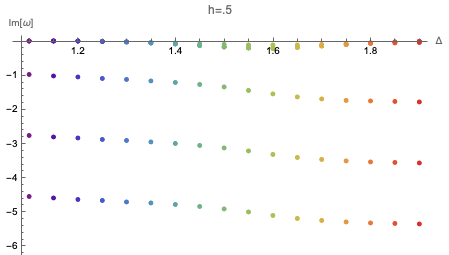}
    \includegraphics[width=0.33\linewidth]{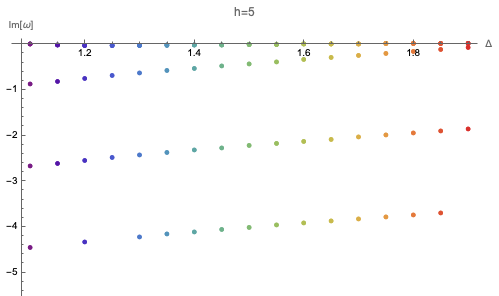}
    \caption{Left: Imaginary component of the modes vs $\Delta$ for a Schwarzschild BTZ black hole with no coupling. Middle: Imaginary part of the modes vs $\Delta$ for a Schwarzschild black hole in AdS coupled to an empty AdS with coupling constant $h=.5$. Right: AdS Schwarzschild coupled to empty AdS with coupling constant $h=5$. Due to the coupling, there are extra set of modes picked up on the plots of finite $h$. For small h such that $h(2\Delta-2)=g <1$, the imaginary component of the modes has a similar behavior to the no coupling case. However for values of h such that $h(2\Delta-2)=g >1$ we have the imaginary component of the modes becoming less negative. The transition from decreasing into increasing happening approximately at $\Delta = \frac{1}{2h}+1$.}
    \label{fig:diffdel1bh}
\end{figure}

\begin{figure}
    \centering
    \includegraphics[width=0.45\linewidth]{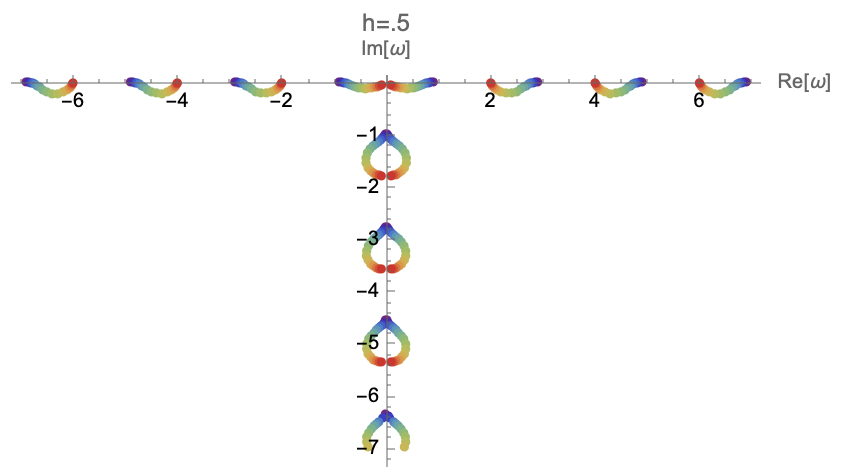}
    \includegraphics[width=0.45\linewidth]{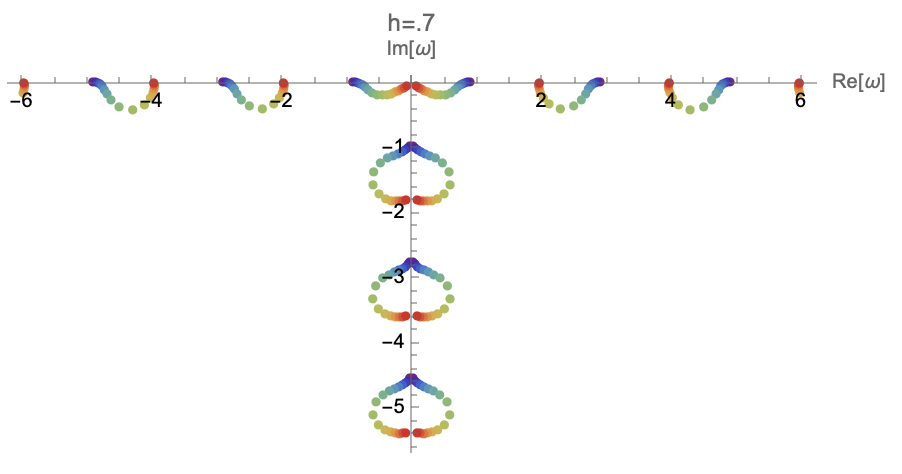}
    \includegraphics[width=0.45\linewidth]{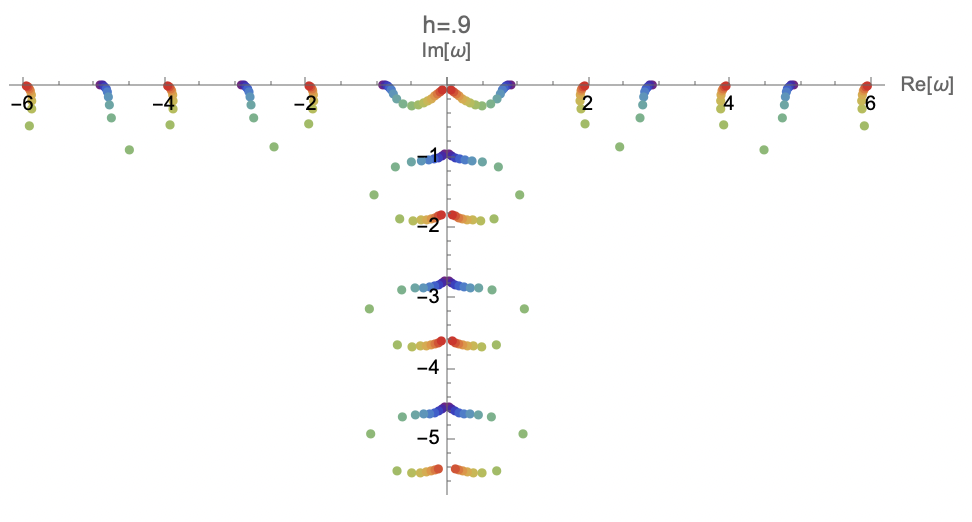}
    \includegraphics[width=.45\linewidth]{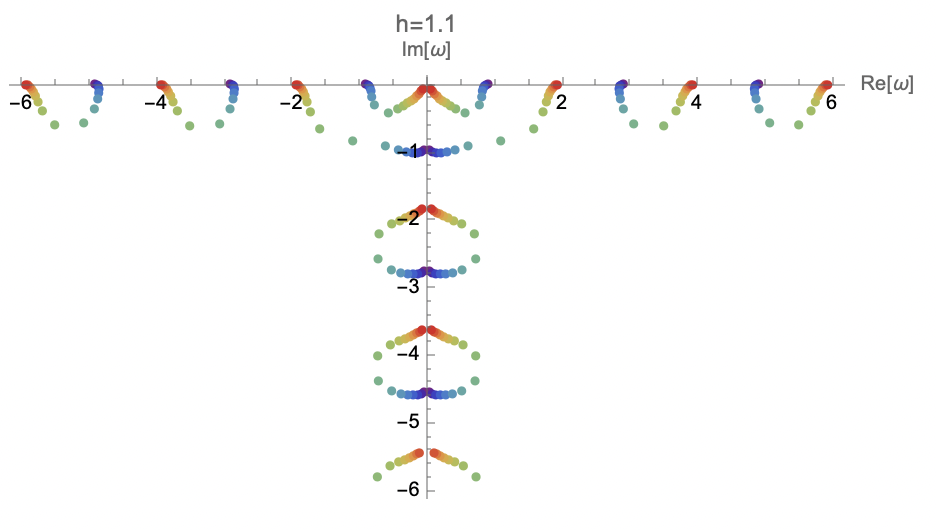}
    \includegraphics[width=.45\linewidth]{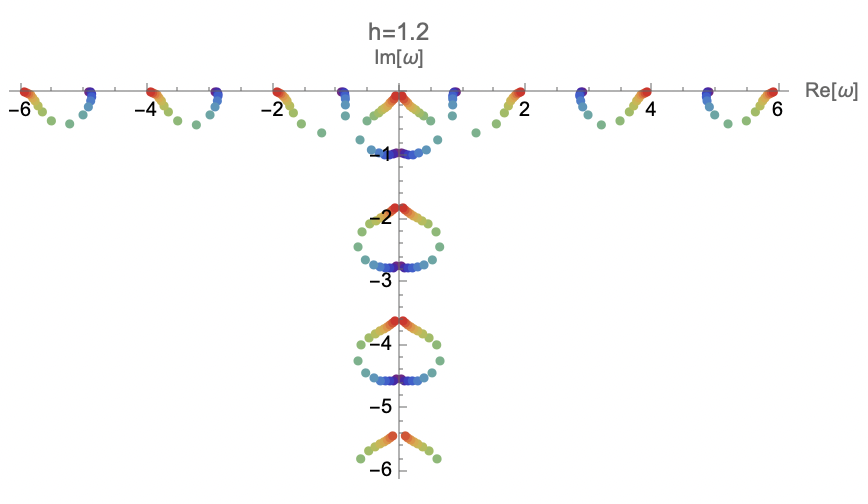}
    \includegraphics[width=.45\linewidth]{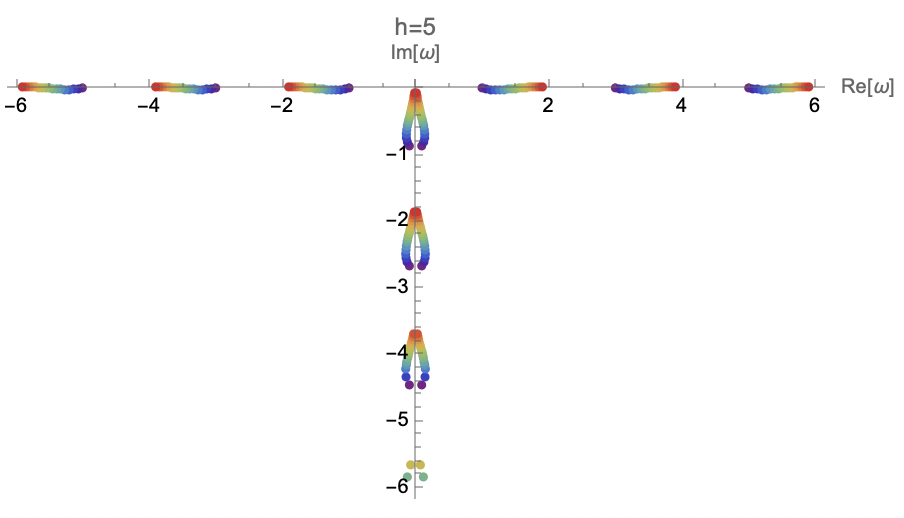}
    
    \caption{Each plot represents a fixed value of $h$. On each plot, we show quasinormal modes for different values of $\Delta$. Colors indicate $\Delta$, with purple for $\Delta=1.1$, red for $\Delta=1.9$, and intermediate values shown in increments of $.05$.}
    \label{fig:1bhvaryingdelmodeshbelowabove1}
\end{figure}

\begin{figure}
    \centering
    \includegraphics[width=0.4\linewidth]{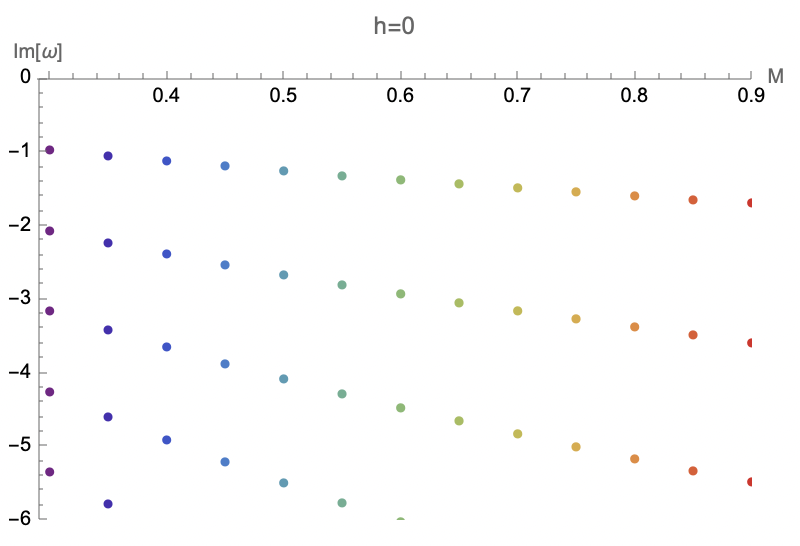}
    \quad
    \includegraphics[width=0.42\linewidth]{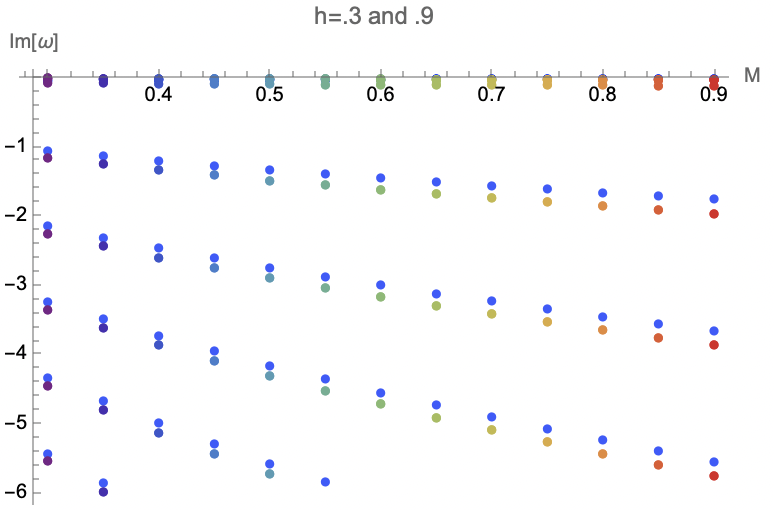}
    \caption{Left: $Im(\omega)$ vs the mass of the black hole for a BTZ without coupling. Right: $Im(\omega)$ vs M for a BH coupled to empty AdS. We use $\Delta=1.8$ and $h=.3$ and $.9$. Due to the coupling, an extra set of modes appear at the top of the right plot. The modes for $h=.3\,(g=.48<1)$ in gradient of blue is slightly less negative than those of $h=.9\, (g=1.44>1)$ in rainbow. The plot shows different values of the black hole mass starting from $.3$ to $.9$ in increments of $.05$. 
    }
    \label{fig:MvsIm1bh}
\end{figure}

In Figure \ref{fig:diffdel1bh}, we see another interesting feature: dissipation. Before coupling the empty AdS has only normal modes, no imaginary component. The black hole however, had only imaginary modes, no real components. After coupling, we see the real modes of AdS starting picking up imaginary components, signaling dissipation. One can see that clearly on Figure \ref{fig:1bhvaryingdelmodeshbelowabove1}. Same dissipative behavior could also be observed in Figure \ref{fig:MvsIm1bh} where we fix $\Delta$ and $h$ and we vary the mass of the black hole. After the coupling, the modes not only pick an extra set that lies close to the real numbers line but also show that as $h$ increases, the modes become slightly more negative.  
 
\subsection{Two Black Holes of Different Sizes} 

\begin{figure}
    \centering
    \includegraphics[width=0.4\linewidth]{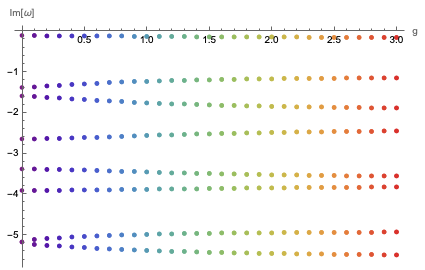} 
    \quad \quad \includegraphics[width=0.05\linewidth]{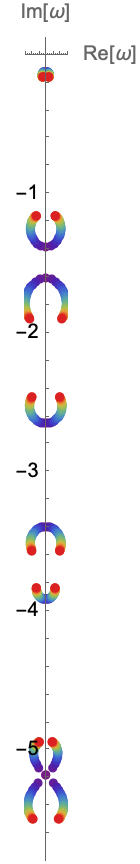} 
    \caption{Two black holes of different sizes with $\Delta=1.8$, $M=.8$ and $M_2=.4$. Values of $g$ go from 0 to 3 in increments of .5. The purple ones on the imaginary axis correspond to what one would expect from setting $\alpha=0$ for a standard quantization of the right black hole and $\beta=0$ for standard quantization of the second black hole.}
    \label{fig:gvsIm2bhs}
\end{figure}

The case of one finite temperature, one zero temperature CFT we just considered is the most interesting from the point of view of studying the onset of dissipation due to coupling to the bath. For completeness, we also studied the case of two non-vanishing but non-equal size black holes. The coefficients for two black holes of mass $M$ (on the right) and $M_2$ (on the left) are:
\begin{equation}
    \begin{split}
\alpha_R &=-\frac{\pi  \csc (\pi  \Delta ) \Gamma \left(1-\frac{i \omega }{\sqrt{M}}\right)}{\Gamma (2-\Delta ) \Gamma \left(\frac{1}{2} \left(\Delta -\frac{i (\omega -m)}{\sqrt{M}}\right)\right) \Gamma \left(\frac{1}{2} \left(\Delta -\frac{i (m+\omega )}{\sqrt{M}}\right)\right)}\\
\beta_R &=\frac{\pi  \csc (\pi  \Delta ) \Gamma \left(1-\frac{i \omega }{\sqrt{M}}\right)}{\Gamma (\Delta ) \Gamma \left(\frac{1}{2} \left(-\Delta -\frac{i (\omega -m)}{\sqrt{M}}+2\right)\right) \Gamma \left(\frac{1}{2} \left(-\Delta -\frac{i (m+\omega )}{\sqrt{M}}+2\right)\right)}\\
\alpha_L &=\frac{\pi \, \csc (\pi \Delta )\, \Gamma \left(1-\frac{i \omega }{\sqrt{M_2}}\right)}{\Gamma (\Delta ) \Gamma \left(\frac{1}{2} \left(-\Delta -\frac{i (\omega -m)}{\sqrt{M_2}}+2\right)\right) \Gamma \left(\frac{1}{2} \left(-\Delta -\frac{i (m+\omega )}{\sqrt{M_2}}+2\right)\right)}\\
\beta_L &=-\frac{\pi  \csc (\pi \Delta )\, \Gamma \left(1-\frac{i \omega }{\sqrt{M_2}}\right)}{\Gamma (2-\Delta ) \Gamma \left(\frac{1}{2} \left(\Delta -\frac{i (\omega -m)}{\sqrt{M_2}}\right)\right) \Gamma \left(\frac{1}{2} \left(\Delta -\frac{i (m+\omega )}{\sqrt{M_2}}\right)\right)}\\
\end{split}
\label{eq:2bhscoeff}
\end{equation}
In a similar fashion to the previous case, the loops on Fig \ref{fig:gvsIm2bhs} show the duality between strong and weak coupling. Larger values of the couplings drives the system from the right black hole in the standard quantization and the left in the alternate to the right black hole in the alternate quantization and the left black hole in the standard.

\begin{figure}
    \centering
    \includegraphics[width=0.45\linewidth]{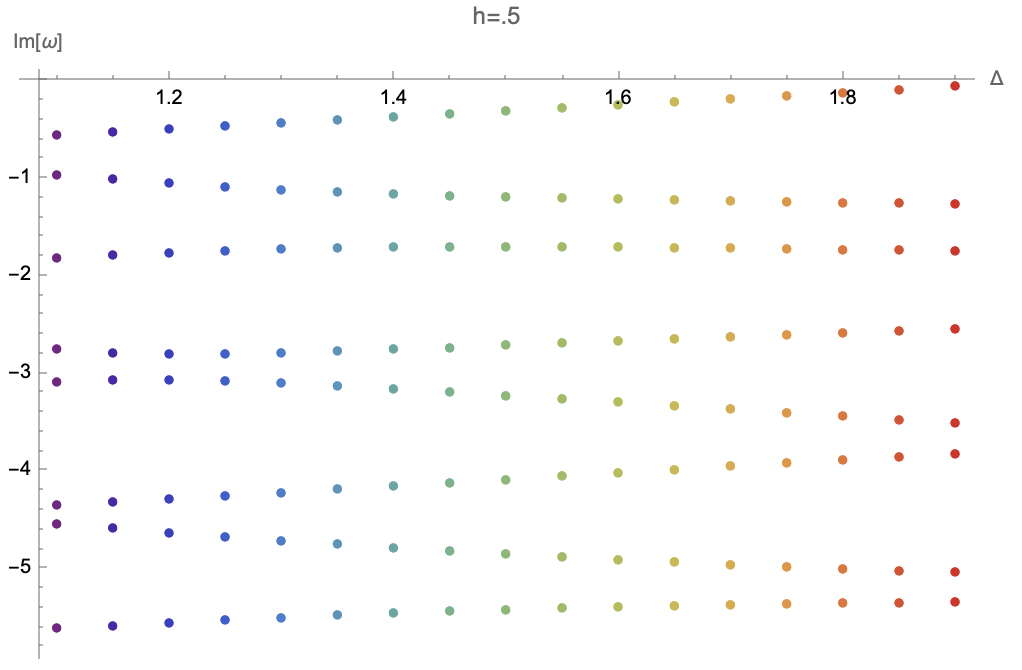}
    \includegraphics[width=0.45\linewidth]{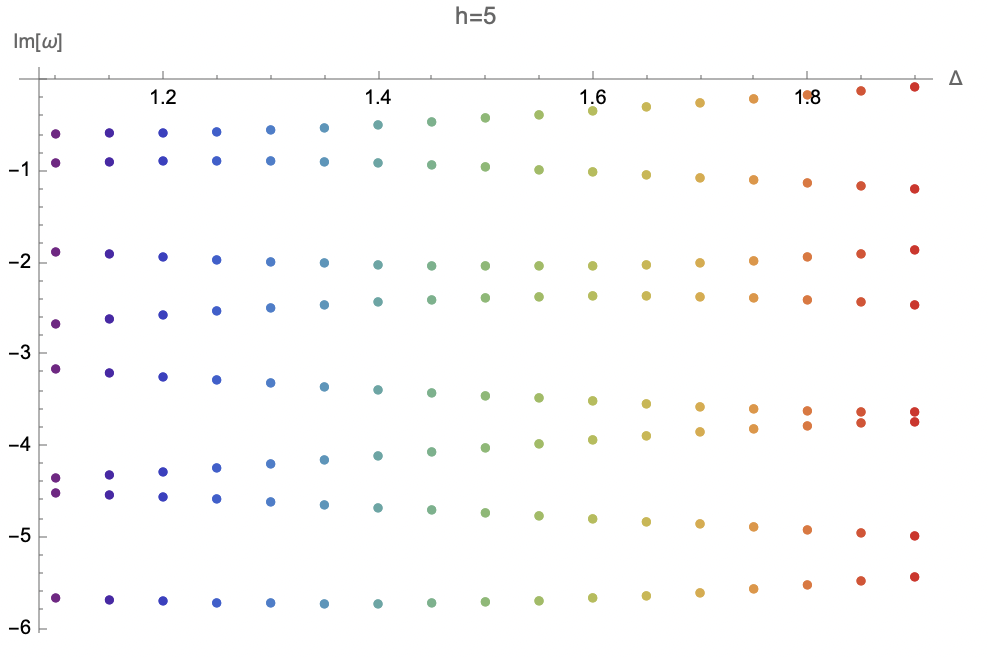}
    \caption{$Im(\omega)$ vs $\Delta$ for different values for $h$. $M$=.8, $M_2$ =.4.}
    \label{fig:delvsIm2bhs}
\end{figure}

\begin{figure}
\centering
\includegraphics[width=0.115\linewidth]{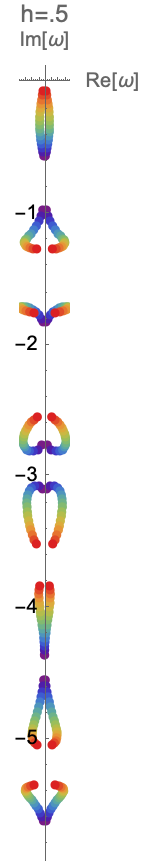}
\includegraphics[width=0.12\linewidth]{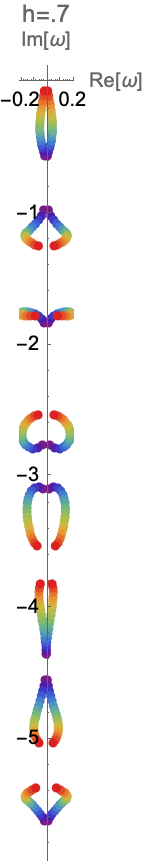}
\includegraphics[width=0.12\linewidth]{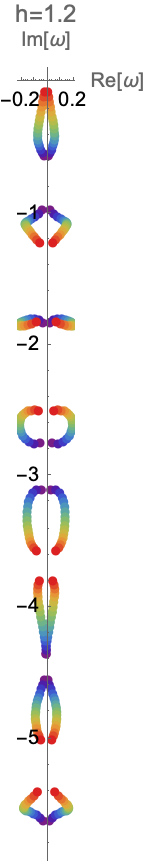}
\includegraphics[width=0.12\linewidth]{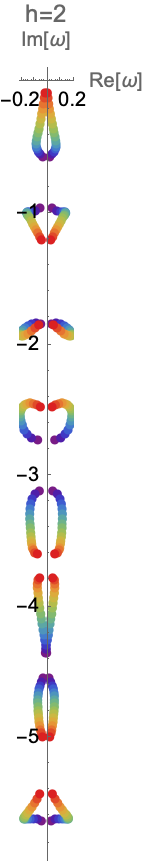}
\includegraphics[width=0.11\linewidth]{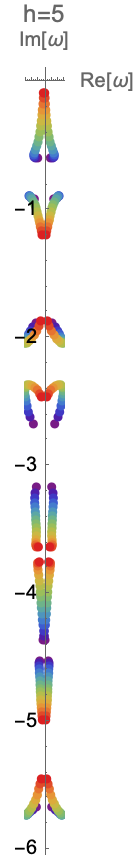}
\includegraphics[width=0.1\linewidth]{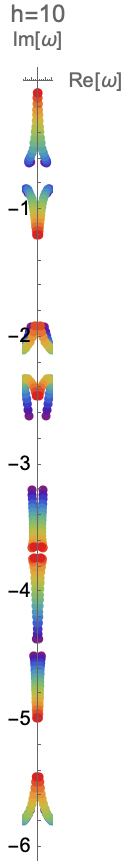}
    \caption{Quasinormal modes for different $\Delta$. Colors indicate $\Delta$ ranging from purple: $\Delta=1.1$ to red: $\Delta=1.9$ in increments of .05. $M=.8 \, \&\,$$M_2 =.4$ for different values of $h$.}   \label{fig:modes2bhsdiffdel}
\end{figure}

\begin{figure}
    \centering \includegraphics[width=0.4\linewidth]{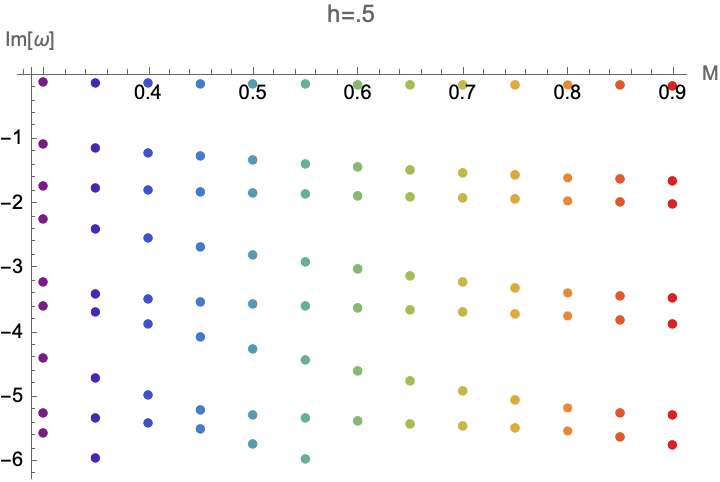}
    \quad  \includegraphics[width=0.4\linewidth]{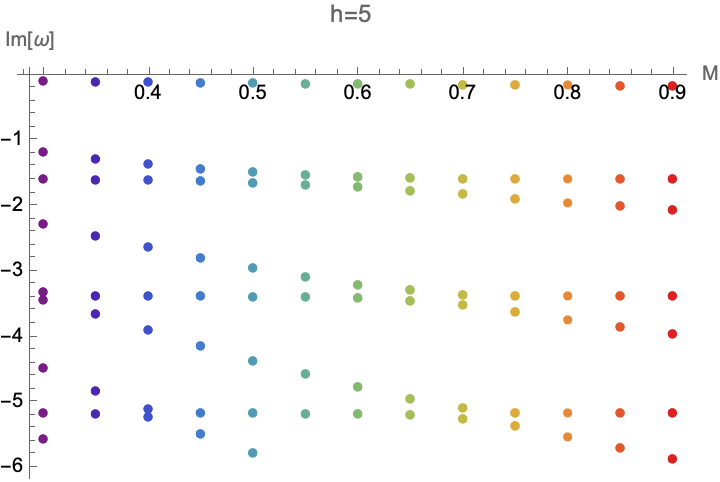}   \includegraphics[width=0.4\linewidth]{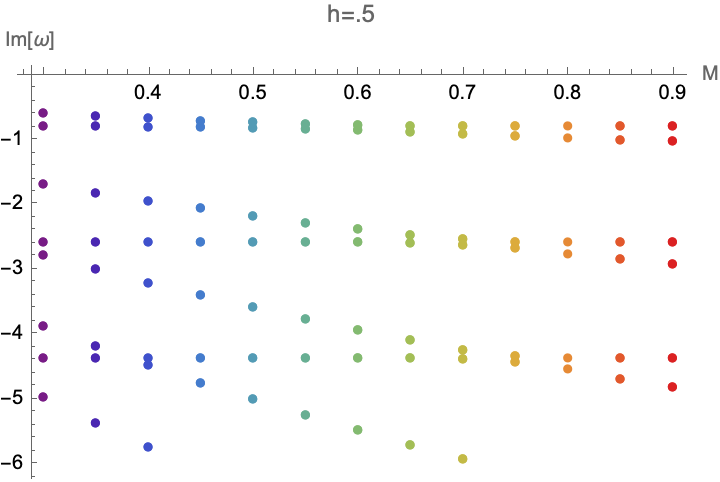}
    \quad \includegraphics[width=0.4\linewidth]{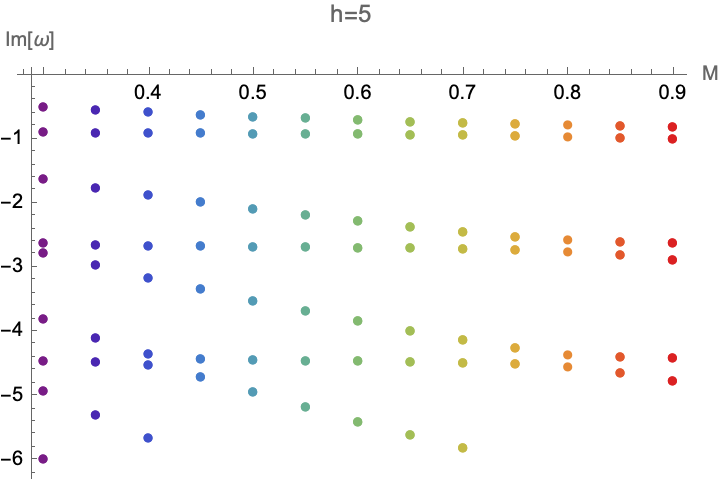}
    \caption{$Im(\omega)$ vs M for different values of $h$. $\Delta=1.8 \to$ top panels and $\Delta=1.1 \to$ bottom panels. $M_2=.8$.}
    \label{fig:varingMfor2bhs}
\end{figure}

While it is not clear how to interpret the interesting behavior we see of Figure \ref{fig:delvsIm2bhs} and Figure \ref{fig:modes2bhsdiffdel} for the two coupled black hole case, dissipation is still clear from \ref{fig:varingMfor2bhs}. Note the spacing between the modes in Figure \ref{fig:delvsIm2bhs} and Figure \ref{fig:varingMfor2bhs} is quite different from the single black hole case in the left of Figure \ref{fig:diffdel1bh} and Figure \ref{fig:MvsIm1bh}. By comparing the top and bottom panels of Figure \ref{fig:varingMfor2bhs}, we notice that as you decrease $\Delta$, the first set of the Imaginary modes becomes more negative. In other words, the rate of decay gets faster. 

\clearpage
\section{Discussion}
In this article, we explored how a CFT and its dual gravitational system behave when coupled to a bath under a double-trace deformation which serves as a useful proxy for studying open-system dynamics. Although our analysis considered the combined system and environment as a whole—hence preserving unitary evolution—we found that the quasi-normal modes of a zero temperature CFT coupled to finite temperature bath exhibit dissipation, a hallmark of open quantum systems. 

In calculating transmission coefficients, we found that they are independent of the bulk geometry. As a sanity check, we tested the conformally coupled scalar case where AdS should map to Minkowski space and we found perfect agreement between the transmission and reflection coefficients of a free scalar in flat space to those of AdS with conformally coupled scalar. We have also found a new strong/weak coupling duality where upon increasing the coupling for a fixed conformal dimension, we drive the system from standard quantization on the right and alternate on the left to standard on the left and alternate on the right as evident in Figure \ref{fig:1bhmodesdiffg} for 1 black hole coupled to empty AdS and Figure \ref{fig:gvsIm2bhs} for 2 coupled black holes.

The quasi-normal modes showed other interesting behavior in the one black hole case when varying the coupling and the conformal dimension as in Figure \ref{fig:1bhvaryingdelmodeshbelowabove1}. We notice that there is crossing between the clusters across different couplings; compare the first plot on the upper left to the one on the bottom right. The two-black hole-case, on the other hand, show a complete different behavior to the no coupling case as depicted in Figure \ref{fig:diffdel1bh} on the far left vs those on Figure \ref{fig:delvsIm2bhs}. As for varying the mass, the one black hole case show that the heavier the mass the stronger the dissipation, as it moves the imaginary component of the modes to be more negative as in Figure \ref{fig:MvsIm1bh}.

Notably, throughout this work, we have kept the bath degrees of freedom. While a system-bath coupling could be evolved unitarily as studied in this manuscript, black hole evaporation reveals non-unitary dynamics at the level of subsystems which could lead to effective Lindblad-type dynamics, offering a route to explore non-Hermitian extensions of holography. These themes resonate strongly with developments in condensed matter physics, where open-system dynamics and the behavior of mixed states are active areas of exploration—particularly in understanding dissipation, decoherence, and emergent phenomena far from equilibrium.

One natural avenue to investigate is to trace over the bath degrees of freedom, leaving the system in a mixed state. An immediate question is how Lindbladian dynamics emerges in the presence of a black hole and what the corresponding boundary dual might be. Can we get a non trivial page curve for the entropy of such theory? \cite{ishii2025lindbladdynamicsholography} established a bulk dual for a general free CFT$_2$ undergoing Lindbladian evolution but noted that the a non-zero entropy arises only after the back reaction on the geometry is included. Black hole evaporation provides an ideal playground for studying open systems and it would be interesting to see how the resulting mixed states influence entanglement entropy and entanglement islands.

\section*{Acknowledgements}
We'd like to thank Massimo Porrati, Mianqi Wang, and Aaron Zimmerman for useful discussions. This work was supported in part by DOE grant DE-SC0022021 and by a grant from the Simons Foundation (Grant 651678, AK).

%\clearpage
%\newpage
%\vfill
%\pagebreak

\bibliographystyle{jhep}
\bibliography{main2}
\end{document}